\documentclass[sigconf]{acmart}
\usepackage{soul}
\usepackage{amsmath}
\usepackage{tabularx}
\usepackage{fontawesome5} 
\usepackage{pifont}
\usepackage{subcaption}
\usepackage[dvipsnames]{xcolor}
\usepackage{algpseudocode}
\usepackage[linesnumbered, ruled, vlined]{algorithm2e}
\usepackage{tikz}
\usepackage{booktabs}
\usepackage[most]{tcolorbox}
\usetikzlibrary{shapes.geometric, arrows, positioning}
\usepackage{enumitem}
\usepackage{adjustbox}
\usepackage{siunitx}
\usepackage{makecell}
\usepackage[table]{xcolor}
\usepackage{longtable,array}
\newcommand{\AgentRole}{\textit{<AgentRole>}}
\usepackage{tabularx}
\usepackage{xcolor}
\definecolor{peach!20}{RGB}{252, 230, 205}
\definecolor{blue!15}{RGB}{217, 226, 243}

\definecolor{novicegreen}{RGB}{210,235,210}   
\definecolor{interblue}{RGB}{200,220,255}     
\definecolor{expertred}{RGB}{255,210,210}     

\newcommand{\participant}[2]{%
  \colorbox{#2!40}{\sffamily #1}%
}
\newcommand{\N}[1]{\participant{#1}{novicegreen}} 
\newcommand{\I}[1]{\participant{#1}{interblue}}   
\newcommand{\E}[1]{\participant{#1}{expertred}} 

\newcolumntype{L}[1]{>{\raggedright\arraybackslash}p{#1}}
\newcolumntype{C}[1]{>{\centering\arraybackslash}p{#1}}
\AtBeginDocument{%
  }

\setcopyright{acmlicensed}
\copyrightyear{2018}
\acmYear{2018}
\acmDOI{XXXXXXX.XXXXXXX}
\acmConference[Conference acronym 'XX]{Make sure to enter the correct
  conference title from your rights confirmation email}{June 03--05,
  2018}{Woodstock, NY}
\acmISBN{978-1-4503-XXXX-X/2018/06}

\newcommand{\cmark}{\textcolor{novicegreen}{\ding{51}}}
\newcommand{\xmark}{\textcolor{expertred}{\ding{55}}}
\definecolor{virt-lab}{HTML}{FFF2CC}
\newcommand{\toolname}{\textsc{VirT-Lab}}

\usepackage{adjustbox}
\usepackage{colortbl}   
\usepackage{xcolor}  
\usepackage{adjustbox}
\usepackage[most]{tcolorbox}
\usepackage{pifont}
\definecolor{novicegreen}{HTML}{77DD77}
\definecolor{expertred}{HTML}{E06666}
\definecolor{interblue}{HTML}{6D9EEB}

\begin{document}




\title{Simulating Teams with LLM Agents: Interactive 2D Environments for Studying Human–AI Dynamics}


 \author{Mohammed Almutairi}
 \affiliation{%
   \institution{University of Notre Dame}
   \city{Notre Dame}
   \state{Indiana}
   \country{USA}
 }

 \author{Charles Chiang}
 \affiliation{%
   \institution{University of Notre Dame}
   \city{Notre Dame}
   \state{Indiana}
   \country{USA}
 }

 \author{Haoze Guo}
 \affiliation{%
   \institution{University of Notre Dame}
   \city{Notre Dame}
   \state{Indiana}
   \country{USA}
 }
 \author{Matthew Belcher}
 \affiliation{%
   \institution{University of Notre Dame}
   \city{Notre Dame}
   \state{Indiana}
   \country{USA}
 }

\author{Nandini Banerjee}
 \affiliation{%
   \institution{University of Notre Dame}
   \city{Notre Dame}
   \state{Indiana}
   \country{USA}
 }
 \author{Maria Milkowski}
 \affiliation{%
   \institution{University of Notre Dame}
   \city{Notre Dame}
   \state{Indiana}
   \country{USA}
 }
 \author{Svitlana Volkova}
 \affiliation{%
   \institution{Aptima, Inc.}
   \city{Woburn}
   \state{Massachusetts}
   \country{USA}
 }

\author{Daniel Nguyen}
 \affiliation{%
   \institution{Aptima, Inc.}
   \city{Woburn}
   \state{Massachusetts}
   \country{USA}
 }

 \author{Tim Weninger}
 \affiliation{%
   \institution{University of Notre Dame}
   \city{Notre Dame}
   \state{Indiana}
   \country{USA}
 }

 \author{Michael Yankoski}
 \affiliation{%
   \institution{William and Mary}
   \city{Williamsburg}
   \state{Virginia}
   \country{USA}
 }

 \author{Trenton W. Ford}
 \affiliation{%
   \institution{William and Mary}
   \city{Williamsburg}
   \state{Virginia}
   \country{USA}
 }

 \author{Diego G\'omez-Zar\'a}
 \affiliation{%
   \institution{University of Notre Dame}
   \city{Notre Dame}
   \state{Indiana}
   \country{USA}
 }

\renewcommand{\shortauthors}{Almutairi et al.}

\begin{abstract}
  Enabling users to create their own simulations offers a powerful way to study team dynamics and performance. We introduce \toolname{}, a system that allows researchers and practitioners to design interactive, customizable simulations of team dynamics with LLM-based agents situated in 2D spatial environments. Unlike prior frameworks that restrict scenarios to predefined or static tasks, our approach enables users to build scenarios, assign roles, and observe how agents coordinate, move, and adapt over time. By bridging team cognition behaviors with scalable agent-based modeling, our system provides a testbed for investigating how environments influence coordination, collaboration, and emergent team behaviors. We demonstrate its utility by aligning simulated outcomes with empirical evaluations and a user study, underscoring the importance of customizable environments for advancing research on multi-agent simulations. This work contributes to making simulations accessible to both technical and non-technical users, supporting the design, execution, and analysis of complex multi-agent experiments.
\end{abstract}




\begin{teaserfigure}
\centering
  \includegraphics[width=\textwidth]{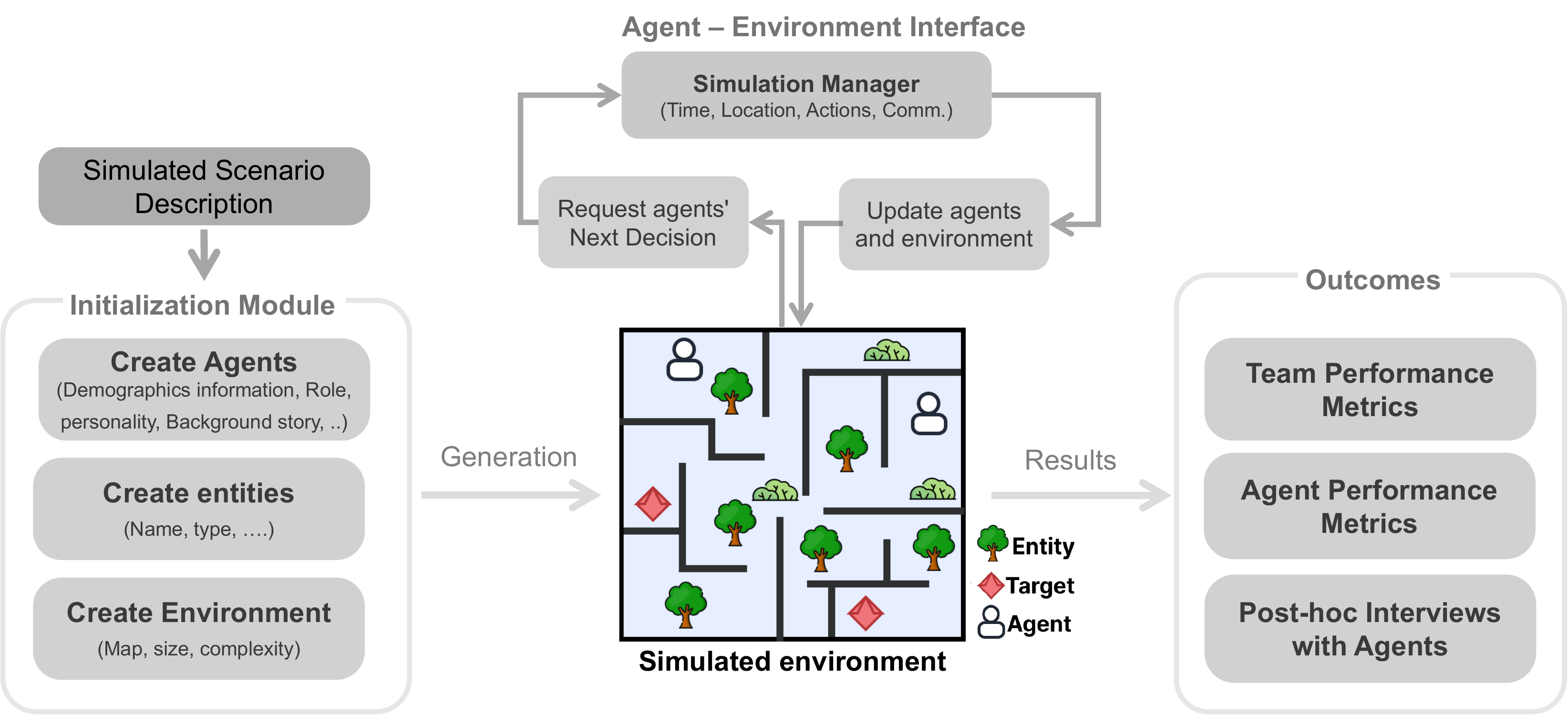}
    \caption{%
    Overview of \toolname{}, an LLM-based multi-agent system for simulating human team behavior. Users define scenarios through natural language instructions, configure agent attributes, and set up the environment. Agents interact autonomously with the environment and each other, generating actions and conversations. The system concludes with structured outputs, including team and agent performance statistics, logs of agent-environment interactions, and a post-simulation interview to support further analysis.
      }
  \label{fig:teaser}
\end{teaserfigure}

\received{20 February 2007}
\received[revised]{12 March 2009}
\received[accepted]{5 June 2009}

\maketitle

\section{Introduction} 
\label{Introduction}

Teams are central to solving society's most complex problems, from disaster response to healthcare and scientific discovery \cite{taylor2025rapidly,wuchty2007increasing,Benishek2019}. Understanding how teams collaborate, adapt, and achieve shared goals has long been a priority for researchers in organizational science and human-computer interaction \cite{mathieu2017century,hong2016human, humphrey2014team, gomez2020taxonomy}. Yet, studying real-world teams in action is difficult: access to real teams is hard, field settings are unpredictable, experiments are costly, testing counterfactual factors is unrealistic, and ethical constraints often limit what researchers can observe \cite{Travis2021,Salas2025,Travis2025}. Simulations offer an alternative by recreating team dynamics in safe and controlled environments, allowing researchers to evaluate hypotheses and conditions that are otherwise infeasible to test in real life. Yet, building simulations that capture the nuance of real teamwork (e.g., communication patterns, shifting roles, relationships) remains a major challenge. Traditional approaches, such as agent-based models \cite{crowder2012development,hsu2016understanding} and reinforcement learning \cite{lapp2019kaboom}, have encoded team behavior through hard-coded rules, which often oversimplify the adaptive nature of human teams \cite{gao2024large}.

Recent advances in multi-agent systems based on Large Language Models (LLMs) present a promising alternative for simulating complex team behaviors. Unlike rule-based approaches, LLMs can generate flexible, context-sensitive responses and adapt their actions to evolving situations. This capacity has motivated more researchers to use multiple LLM-based agents to collaborate within shared environments, emulating aspects of teamwork such as coordination, negotiation, and decision-making \cite{park2024generative, li2023metaagents, park2023generative, chen2023agentverse}. While these systems demonstrate the potential of LLMs to model richer forms of team dynamics, they remain limited in important ways. First, many existing LLM-based frameworks are designed for narrow, task-specific applications, with agents designed to focus on specific task execution (e.g., web development \cite{he2025llm, lu2025uxagent, lee2025learning} and conversational tasks \cite{almutairi2025taifa, ghosh2025yes}). This particular design cannot be easily extended to other simulation contexts and goals, limiting its generalizability and customization. Second, agents in many current frameworks are typically assigned fixed roles, limiting their ability to adjust their behavior or responsibilities in dynamic simulated environments \cite{hong2023metagpt}. Moreover, existing frameworks do not allow users to create and personalize team scenarios embedded in spatial, temporal environments. They are based on static team structures in which agents follow predefined linear workflows \cite{wu2023autogen, pan2024agentcoord}. Lastly, the lack of intuitive user interfaces and comprehensive visualization tools, coupled with the requirement for significant technical expertise to run large-scale simulations efficiently, limits accessibility and practical utility. These cascade rigid designs may not fully contemplate several team dynamics in real-world scenarios, where team members must adjust their priorities in response to evolving circumstances.

To address these gaps, we introduce \textbf{Vir}tual \textbf{T}eaming \textbf{Lab}oratory (\toolname{}), a novel system that leverages LLMs to simulate human team behavior in realistic environments without requiring users for extensive technical expertise. At the core of \toolname{} is a user-friendly interface that enables researchers and practitioners to create team simulations through natural language instructions. Using text prompts, the system assists users in defining the simulation by specifying the scenario, team members, tasks, and surrounding environment. The interface supports iterative refinement, allowing users to provide more details of the members (e.g., personality, initial backstories, memories, and domain knowledge) and the environment (e.g., entities and spatial layout). In the back-end, \toolname{}'s simulation engine translates these descriptions into an initial state, populating agents, physical entities, and spatial layouts, and manages the dynamic interactions among them. Each LLM agent is initialized as an autonomous entity capable of perceiving and responding to changes in its environment, enabling simulations that evolve in complex and realistic ways. 

Once the simulation settings are established by the user, \toolname{} starts the simulation and enables the LLM agents to interact autonomously with the environment and each other, generating events aligned with their objectives. These events are managed and coordinated by an event scheduling system, ensuring that events are executed in the appropriate sequence and facilitating coordination between LLM agents in evolving scenarios. In addition, \toolname{} supports communication between LLM agents through dyadic and multi-agent conversations that influence each agent's internal state, including short-term and long-term memories, to better emulate realistic team dynamics. The simulation is displayed in a 2D map, where the user can monitor the progression of the simulation, including agents' rationale, conversations, and actions (Figure \ref{fig:teaser}). The \toolname{} concludes the simulation when the specified simulation duration ends or the team's objective is achieved. At the end of the simulation, \toolname{} provides a structured output, including metrics to study teams' actions, a detailed log of agent-environment interactions, and statistical evaluations of both the overall simulation and the performance of individual LLM agents, allowing further analysis.

We demonstrate \toolname{}'s efficacy through qualitative and quantitative evaluations: (1) a controlled evaluation involving systematic ablation studies to measure the importance of each component; (2) an ground truth evaluation simulating a rescue mission scenario conducted with real teams; (3) a user study with 12 participants, including professional experienced in simulations, evaluating the system’s usability, interpretability, and perceived realism; and (4) a numerical study running the system across multiple scenarios with increasing agent numbers and varying environment complexity to evaluate performance under different simulation scales. Across these evaluations, we find that the \toolname{}'s architecture provides the necessary realism to simulate team dynamics in a customizable, intuitive way. 

In sum, this paper makes the following contributions: 
\begin{enumerate}
    \item We introduce \toolname{}, a novel system that leverages LLMs to simulate team behaviors through configurable multi-agent systems within dynamic environments.
    \item We provide an open-source implementation available at [Anonymized Link], offering the research community a versatile tool to study team dynamics in varied scenarios and environments. 
    \item We conduct a comprehensive set of evaluations to assess \toolname{}'s realism, validity, and usability; demonstrating the critical contributions of the architectural components to overall simulation effectiveness. 
    \item We discuss potential future research directions and the implications of using LLM multi-agent simulations, highlighting both opportunities and considerations for responsible usage in studying human team behaviors.
\end{enumerate}

\section{Related Work}
\label{Related_work}

\subsection{Agent-based Modeling for Simulating Human Teams} 
Simulation provides a means of replicating real-world phenomena through computational models \cite{de2019mesoscopic, alluhaybi2019survey}. It has been widely used to study human teams, offering insights into the dynamics of team performance and decision-making processes \cite{wang2018agent}. A typical simulation framework consists of three components: \textit{agents}, which are defined by their attributes and capabilities to interact with their surroundings; the \textit{environment}, where agents perform actions; and the \textit{interaction rules}, which specify how agents relate to each other and to the environment \cite{mili2006divas, macal2005tutorial}. The design of these components varies depending on the simulation approach. Among the most widely used approaches to model human behavior in teams are Agent-Based Modeling (ABM) and Reinforcement Learning (RL) \cite{nguyen2020deep}. 

On one hand, ABMs simulate collective behavior by defining agents that follow predetermined rules for decision-making and interactions. For example, Lapp et al. \cite{lapp2019kaboom} developed the KABOOM framework, which modeled team cognitive styles in problem solving. Teams were formed based on their cognitive style to analyze problem-solving, where agents explore and exploit different possibilities to maximize a scalar objective function to optimize performance. Similarly, Secchi and Neumann \cite{secchi2016agent} discuss how ABMs can be used to enhance the study of social agency and organizational behavior, highlighting the relative ease of programming and ability to borrow simulation practices from other disciplines. While ABMs can effectively capture generalizable interaction patterns in collective environments, they lack the flexibility to represent the nuanced, adaptive, and emergent qualities of real-world social interactions \cite{rand2021agent,an2021challenges}. Because they rely on pre-coded behavioral rules, ABMs often struggle to simulate dynamics, such as communication or trust, and are frequently tailored to domain-specific tasks, which restricts their applicability across broader contexts \cite{boero2005,miller2015agent}.

On the other hand, RL provides a more adaptive alternative by enabling agents to learn from their environment through trial and error. RL has been used to simulate aspects of human teaming, such as trust and ethics \cite{meimandi2023rl}, cooperative problem solving \cite{littman2015reinforcement}, and decision-making under high variability \cite{lake2017building}. For example, Meimandi et al. \cite{meimandi2023rl} developed an RL-based framework that incorporated beliefs, prior knowledge, and social norms into agents' decision processes, helping test hypotheses related to trust across agents. Leibo et al. \cite{leibo2017multi} showed how multi-agent RL can model cooperative and competitive strategies in shared environments with high realism. These approaches illustrate RL's potential to capture adaptive and context-dependent behavior. However, RL requires extensive training, can be computationally expensive, requires advanced skills, and often depends on fixed parameters that may oversimplify the complexities of human cognition and decision making \cite{gao2024large, zhang2021synergistic, littman2015reinforcement, lake2017building}. Moreover, RL frameworks generally focus on optimizing task performance and coordination efficiency, without accounting for language use, communication breakdowns, or other complex social dynamics that shape real-world team behavior \cite{nguyen2020deep,rahwan2019machine}.

These challenges highlight the need for simulation frameworks that can capture the complexity of team dynamics while remaining adaptable and easy to use. \toolname{} takes inspiration from ABM and RL designs by providing a scalable, customizable platform that enables researchers to design, run, and evaluate team simulations.

\subsection{LLM-based Agents for Team Simulation} 
LLMs offer greater flexibility for simulating complex social interactions and evolving agent states than traditional modeling approaches \cite{gao2024large}. Recent studies have shown that LLM-based agents can generate, decompose, and adapt action plans, producing behaviors that appear more human-like and contextually appropriate given the inputs \cite{Song2024, guo2024embodied, zhang2023building, wu2023autogen}. For instance, Park et al. \cite{park2023generative} demonstrated that LLM-based agents could simulate believable individual routines, conversations, and decisions that adapt dynamically to changes in their environments. Similarly, Fan et al. \cite{fan2024comp} showed that LLM-based agents can retain relevant memories, act consistently over time, and engage in socially grounded behaviors, including forming relationships, recalling previous experiences, and coordinating group activities. 

Most existing LLM-based agent frameworks approach collaboration primarily through conversational exchanges. A common strategy is to design agents that communicate with one another to solve problems, often by assigning them specific roles \cite{guo2024large}. For example, Wu et al. \cite{wu2023autogen} introduced ``AutoGen,'' a framework for building customizable role-specific agents that coordinate through structured dialogues. A central LLM coordinates the agents' conversations, deciding who should speak next. Building on this approach, Song et al. \cite{Song2024} presented a hierarchical collaboration framework in which a ``builder agent'' decomposes a complex task, recruits other agents with appropriate roles and tools, and coordinates their efforts. In addition, a ``reflector agent'' evaluates the team’s performance and prompts adjustments to improve collaboration. These systems demonstrate the potential of conversational coordination. 

Recent work has begun to integrate spatial information and embodied interactions into LLM-based simulations. Zhang et al. \cite{zhang2023building} developed a cooperative multi-agent system in which agents could perceive their surroundings, update internal memory, communicate, and collaboratively plan and execute tasks. However, this approach was limited to two agents and struggled with integrating more complex spatial reasoning and environmental constraints. In a more recent effort, Guo et al. \cite{guo2024embodied} extended the AutoGen framework to support multi-agent coordination in physical environments involving three or more agents. These frameworks and many others represent a promising step toward simulating dynamic, goal-driven, and socially coherent multi-agent teams \cite{pan2024agentcoord, park2024generative, zhou2023sotopia}. The next challenge is to assess the extent to which such simulated team dynamics align with real-world team coordination and collaboration.

Despite advances in spatial and conversational coordination, most existing frameworks remain challenging for non-experts to adopt. They often lack intuitive user-facing tools such as web-based interfaces or flexible scenario editors, limiting accessibility for practitioners without specialized technical skills. A few examples are ``SOTOPIA-S4'' \cite{zhou-etal-2025-sotopia}, which offers a simulation engine and web interface for designing multi-party LLM-based social conversations without programming. Similarly, ``AgentScope-Manager'' supports large-scale multi-agent simulations via a web interface that enables monitoring, agent configuration, and background generation across devices \cite{gao2024agentscopeflexiblerobustmultiagent}. Despite these advances, these web-interfaces do not consider spatial movement for these agents. Lastly, while the Generative Agents framework \cite{park2023generative} or ``AgentCoord'' \cite{pan2024agentcoord} include a web interface and multi-party scenarios, it is limited to specific maps designed for these systems. 

\subsection{Evaluating Multi-Agent Systems and Team Behaviors.}
A growing body of research has explored how to assess LLM-based multi-agent systems, from benchmarks in cooperation and coordination to metrics for emergent social behavior\cite{liu2023agentbench,chen2023agentverse,lin2023agentsims, xu2024theagentcompany, tran2025multi}. These efforts highlight the difficulty of evaluating not only individual agent performance but also collective outcomes such as coordination, efficiency, and robustness. Some studies have introduced multi-agent metrics---such as communication quality, planning, and cooperation scores---to assess how well agents work together beyond task completion \cite{liu2025proactiveeval, zhou2023sotopia}. Others have suggested specialized metrics like role differentiation scores to quantify the specific contributions of each LLM-based agent \cite{lu2024morphagent}. Despite these advances, systematic evaluation of complex team dynamics remains limited, with most frameworks providing only partial coverage of the processes that sustain coordination and collaboration. 

To address these gaps, \toolname{} provides a system that supports simulations within spatial environments, multi-party interactions, and a web-based user interface to facilitate the creation of customizable scenarios (see Table \ref{tab:framework-comparison}). It also integrates configurable evaluation metrics directly into the simulation pipeline, enabling users to analyze both process and outcome variables in team dynamics. In addition, we complement these automated metrics with a user study to better understand how researchers design, interact with, and interpret multi-agent simulations.

\begin{table*}[!htb]
  \centering
  \caption{A comparison between \toolname{} and existing state-of-the-art LLM-based multi-agent simulation frameworks.}
  \resizebox{0.9\textwidth}{!}{%
    \rowcolors{2}{white}{gray!10}%
    \begin{tabular}{>{\bfseries}l c c c c c c c c}
      Framework
        & Web-UI
        & Customized Sim Env.
        & Customized Sim Scenario
        & Task-Solving
        & Spacial Managment\\
      \toprule
      OASIS \cite{yang2024oasis}
        &  \xmark  &  \xmark  &  \cmark  &  \xmark  &  \xmark \\
      Generative Agent \cite{park2023generative}
        &  \cmark  &  \xmark  &  \cmark  & \xmark  &  \cmark \\
      AutoGen \cite{wu2023autogen}
        &  \xmark  &  \xmark  &  \cmark  &  \cmark  &  \xmark \\
      MindAgent \cite{gong-etal-2024-mindagent}
        &  \xmark  &  \xmark  &  \xmark  &  \cmark  &  \cmark  \\
      AgentCoord \cite{pan2024agentcoord}
        &  \cmark  &  \cmark  &  \xmark  &  \xmark  &  \xmark \\
      AgentSociety \cite{piao2025agentsociety}
        &  \xmark  &  \xmark  &  \cmark  &  \xmark  &  \xmark \\
      \rowcolor{virt-lab}
      \toolname{} (Ours)
        &  \cmark  &  \cmark  &  \cmark  &  \cmark  &  \cmark \\
      \bottomrule
    \end{tabular}
  }
  \label{tab:framework-comparison}
\end{table*}
\section{\toolname{} System} 
\label{System Description}
Our motivation for designing the \toolname{} system comes from the limitations of existing frameworks, which often require significant technical expertise and computational setup, and offer narrow support for simulating team dynamics. To address these gaps, we developed \toolname{} as a configurable, web-based system. Through a prompting interface, users can create scenarios, define team structures, and observe how autonomous agents collaborate, make decisions, and adapt as the scenarios unfold. By lowering barriers to simulation design while maintaining realism and control, \toolname{} provides a realistic and controlled simulation environment for studying and analyzing complex team behaviors.

\subsection{Design Goals}
\toolname{} system integrates four key design goals informed by our experience in team modeling and simulation, as well as findings from a semi-systematic review of the literature. Following the approach proposed by Snyder\cite{snyder2019literature}, we review the work in three key areas: human dynamics and behaviors \cite{ilgen2005teams, peeters2006personality, mao2016experimental,espinosa2015temporal,muric2019collaboration}, agent-based modeling and reinforcement learning simulations \cite{lake2017building, lapp2019kaboom,rand2021agent, meimandi2023rl}, and LLM-based multi-agent frameworks \cite{wu2023autogen, park2023generative, Song2024,guo2024embodied,fan2024comp}. We articulate our design goals as follows:

\paragraph{\textbf{\textit{DG1: Support cohesive and sustainable simulations.}}} Coordinating multiple autonomous LLM agents introduces several challenges. Agents struggle to model social constructs, such as individual behaviors, collective dynamics, norms, and organizational structures \cite{gurcan2024llm}. When interacting with each other, they may produce hallucinations or inconsistent reasoning that disrupt simulations \cite{sun2024towards}, and they frequently fail to maintain coherent roles or stable interactions over time \cite{li2023metaagents}. These limitations highlight the need for simulation tools that promote stable, realistic, and cohesive interactions in multi-agent environment.
    
\paragraph{\textbf{\textit{DG2: Preserve user agency in simulations.}}} Maintaining user agency is essential in simulation tools. A system should address this by giving granular control over agent characteristics, task specifications, and environmental parameters. Enabling these characteristics requires providing the necessary tools for users to write, edit, and manage scenarios as they evolve. These features will help align simulations with user goals  \cite{an2021challenges, gao2024large}.
    
\paragraph{\textbf{\textit{DG3: Support visual exploration of LLM agent team interactions.}}} Visualizations enable users to understand agent behaviors and team dynamics by transforming overwhelming text-based interactions into clear, insightful visual representations. \toolname{} should provide interactive and dynamic visualization tools to explore team structures, track simulation progress within spatial environments, and analyze both overall simulation metrics and agent-specific behaviors. This capability addresses key challenges identified in prior work related to evolving dynamics between LLM-based agents, where users must iteratively read, interpret, and debug text-based agent behaviors to adjust interactions across the team \cite{takagi2025framework, epperson2025interactive}.

\paragraph{\textbf{\textit{DG4: Lower the Barrier for non-expert users.}}} While LLMs offer flexible and natural interaction paradigms, existing studies show that non-AI experts often struggle to effectively use these systems due to limited prompt literacy, opportunistic trial-and-error behavior, and expectations shaped by human-to-human communication norms \cite{nam2024using, zamfirescu2023johnny, ma2023}. To make LLM-based simulations more accessible and effective for users, \toolname{} should provide structured guidance and affordances that support the design of team simulations with LLM agents. This includes support for defining the simulated environment, agent skills, and personalities, providing clear input to guide the simulation, and interpreting simulation outcomes.

\subsection{Example Scenario}
\label{Example User Scenario}
In this section, we provide an illustrative example of how \toolname{} can be used in practice. Imagine Nikki, a first responder team leader, who wants to optimize team composition for a rescue operation. Using \toolname{}, Nikki develops a search-and-rescue mission scenario taking place in a multi-room facility. She begins by specifying the mission objectives (e.g., \textit{``locating and rescuing two missing victims''}) along with key parameters of the facility layout (e.g., \textit{``...comprising six interconnected rooms with a safe area.''}). Nikki also sets the maximum duration allowed for the simulation, and the success criteria that both missing individuals must be located and safely evacuated to the designated safe area. 

Next, Nikki assembles the response team by instantiating four LLM agents as team members: two searchers, one medic, and one coordinator. The system allows Nikki to customize each agent, giving options to personalize the demographic profiles, personality traits, relevant skills, and background stories. For example, Nikki configures the medic with high conscientiousness, while the searchers vary in communication style and prior experience. She also refines the environment by adjusting spatial complexity, such as modifying the layout of the search area to make the scenario more challenging. Once the setup is complete, Nikki starts the simulation.

During the simulation, Nikki monitors progress in real time through a dashboard that displays an event log, agent conversations, and a 2D map representing the agents' activity. After the mission concludes, \toolname{} generates a detailed report summarizing agent actions, communication patterns, and key performance metrics, including mission duration, area coverage, and number of rescued victims. To examine the agents' experiences in detail, Nikki can also review post-simulation interviews with the agents, which provide insights into the agents' reasoning processes, decisions, and teamwork dynamics.

\begin{figure*}[!ht]
  \centering
  \includegraphics[width=\textwidth]{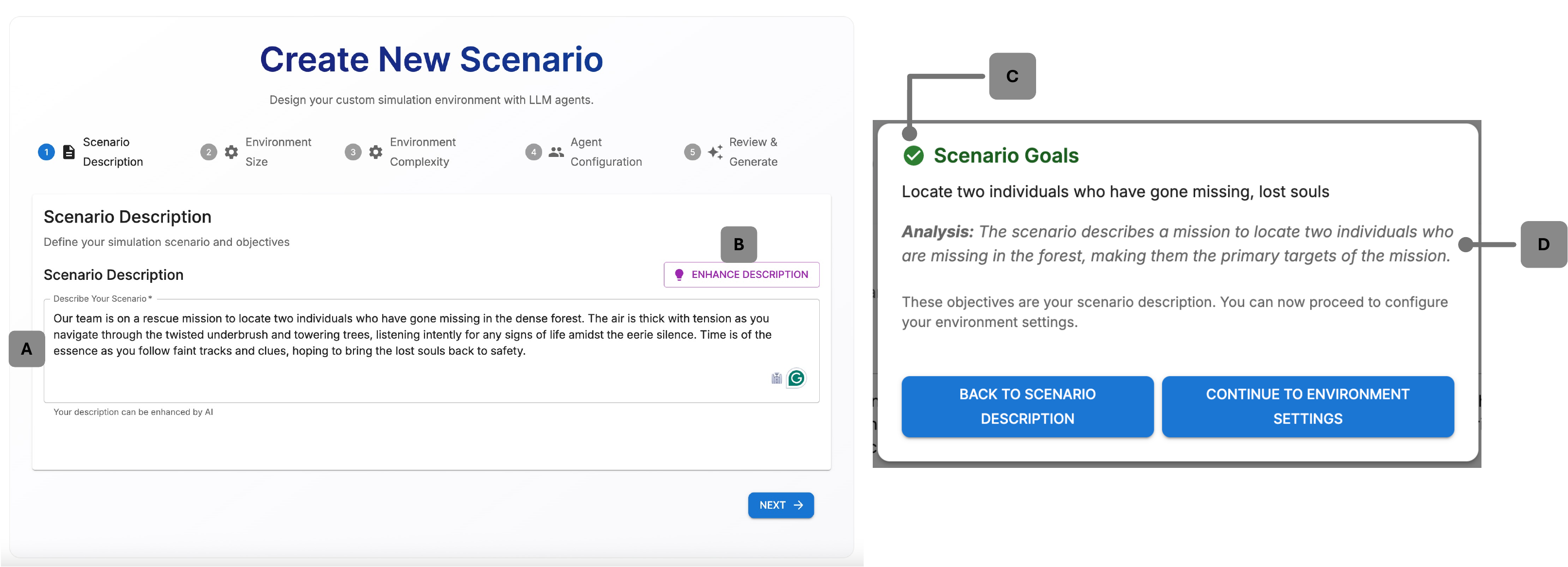}

\caption{Scenario description interface in the \toolname{} system. 
(A) Users enter the description of the simulation scenario, outlining the mission and its objectives. 
(B) The \textit{Enhance Description} button allows users to refine and enhance their input with AI-generated improvements. 
(C) The system detects clear scenario goals based on the user's scenario description. 
(D) Users can review these validated objectives before proceeding.}
  \label{fig:initlization steps}
\end{figure*}

\subsection{Key Features} 
This section introduces the key features of \toolname{}, as illustrated in the example scenario \ref{Example User Scenario}. These features include intuitive scenario setup, agent customization, interactive environment configuration, real-time simulation visualization, agent reasoning inspection, and simulation results analysis.

\subsubsection{Simulation Setup Stage}
A team simulation in \toolname{} begins with the user creating a \textit{scenario} that will define the team composition, the environment, the team goals, and success metrics.

\paragraph{Scenario setup} \toolname{} prompts the user to describe the scenario to be simulated, including the team details, goals, environment, and the maximum duration of the simulation. A scenario could be described as \textit{``Locate two individuals who have gone missing''}, setting the ``scene'' for the simulation. Using a conversational interface, \toolname{} collects user input and asks follow-up questions to clarify or infer additional details needed to generate scenario elements (Figure \ref{fig:initlization steps}a). The system can also enhance the scenario by clicking on \textit{``Enhance Description''} (Figure \ref{fig:initlization steps}b). A pop-up window appears, and the system analyzes the current description to provide more details and suggest modifications to the user (Figure \ref{fig:initlization steps}c), who can accept or reject (Figure \ref{fig:initlization steps}d). 

\begin{figure*}[!ht]
  \centering
  \begin{subfigure}[t]{0.48\textwidth}
    \centering
    \includegraphics[width=\textwidth]{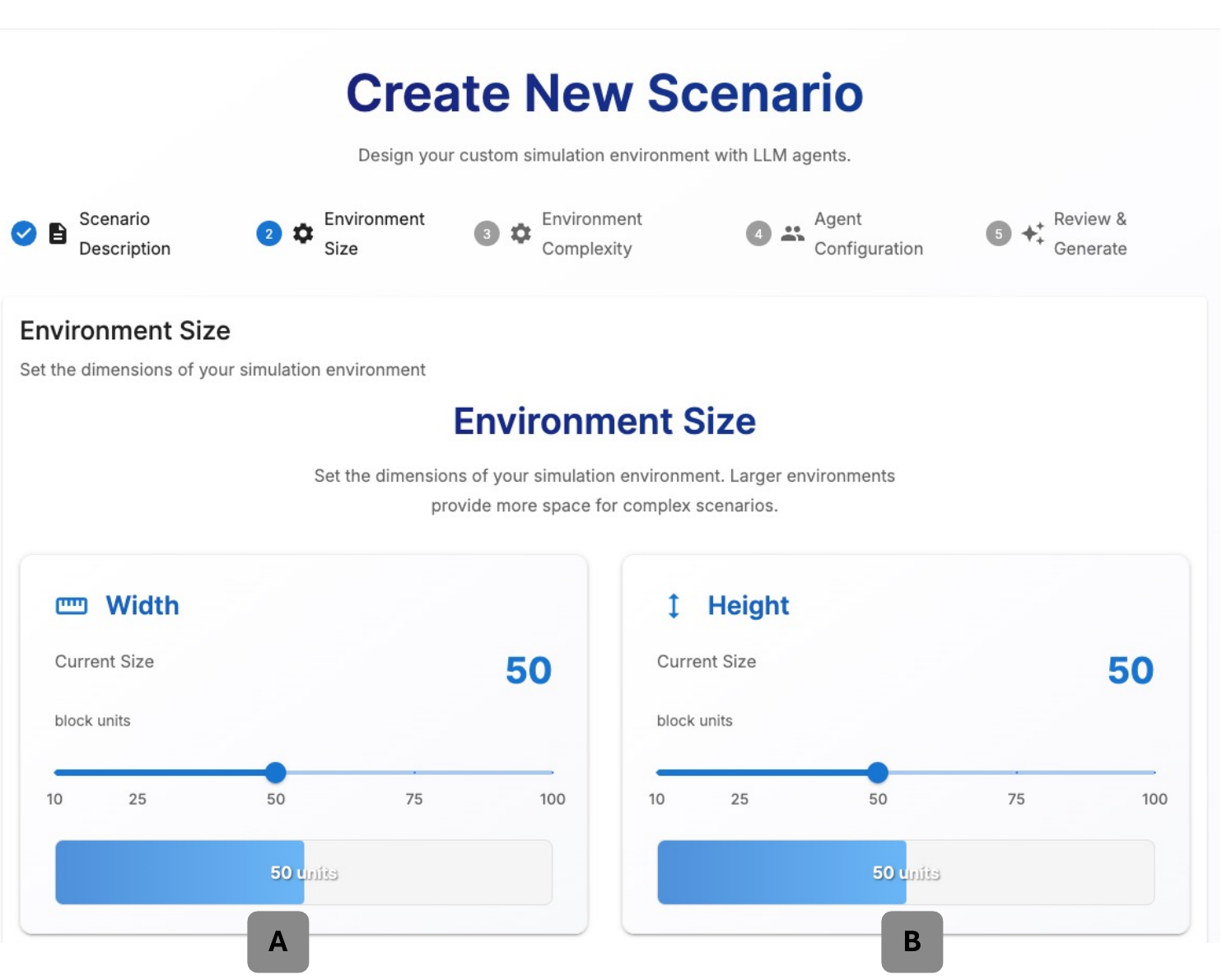}
    \caption{Environment size configuration. Users can adjust the spatial dimensions of the simulation environment by modifying the \textbf{(A)} width and \textbf{(B)} height parameters. Each block unit represents a unit of distance (e.g., a block unit represents a square meter).}
    \label{fig:init_stage}
  \end{subfigure}
  \hfill
  \begin{subfigure}[t]{0.48\textwidth}
    \centering
    \includegraphics[width=\textwidth]{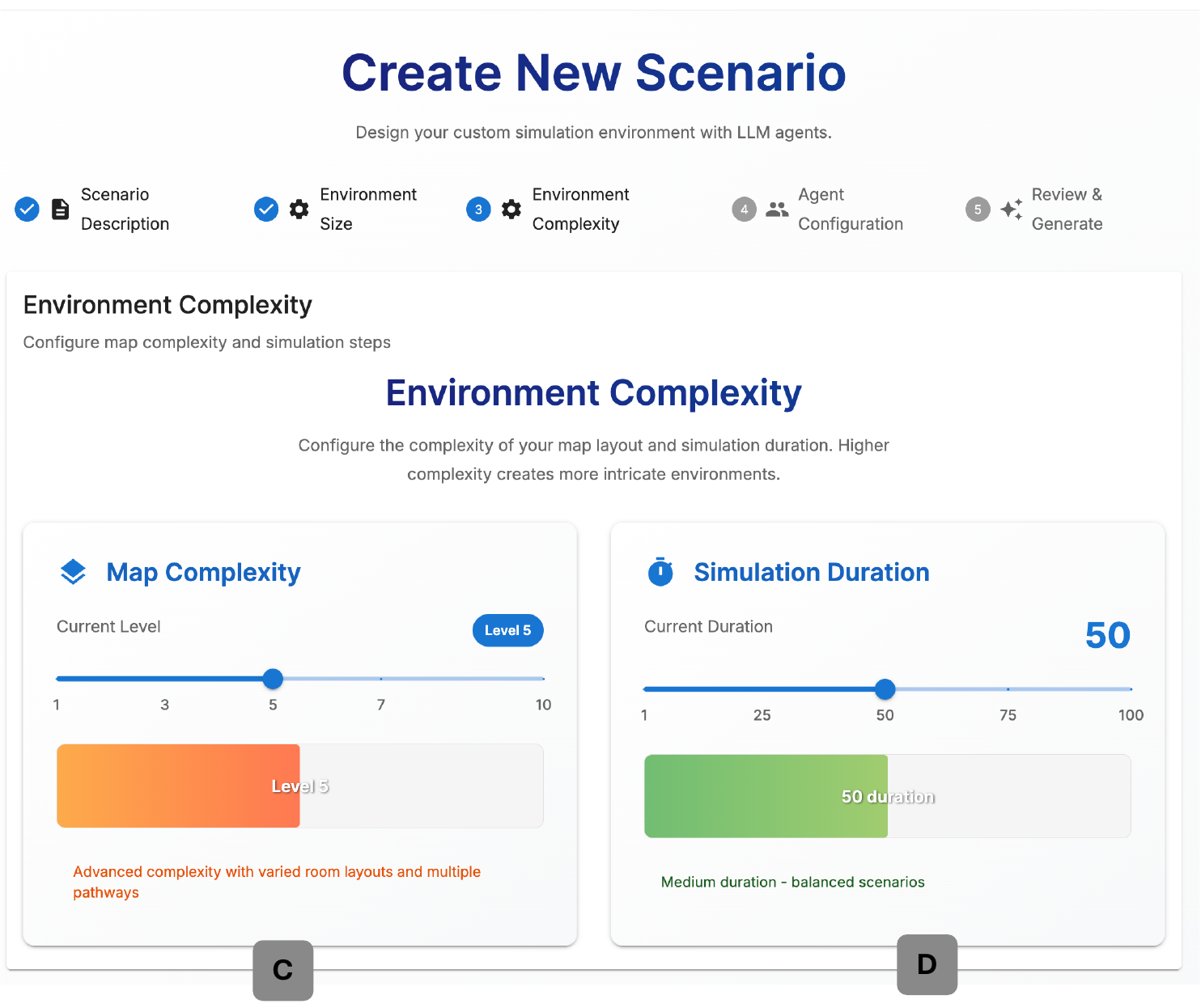}
    \caption{Environment Complexity: users define \textbf{(C)} map complexity and \textbf{(D)} simulation duration to determine the difficulty and length of the scenario}
    \label{fig:env_complex}
  \end{subfigure}
  
\caption{Overview of the \toolname{} interface for simulation setup.}
  \label{fig:env_initlization_combined}
\end{figure*}

The system works with the user to generate a scenario description that includes information on the team composition. For instance, the user might specify \textit{ ``There are two searchers in this scenario.''} and the system will identify two agents to be created. Additionally, the scenario's description might also specify entities relevant to the task (e.g., ``missing victims''), which \toolname{} will create and add to the environment. Finally, the system requires details of the environment, such as environment size (Figure \ref{fig:env_initlization_combined}a) and environment complexity (Figure \ref{fig:env_initlization_combined}b). To evaluate the scenario's success, \toolname{} infers the scenario-specific criteria based on the user-provided description and generates a success function using an LLM. This function is executed at each simulation timestep to determine whether the goals have been met. After these parameters are set, the system displays a refined scenario prompt, allowing the user to iteratively review and modify the scenario's components until confirmation. This flexible, iterative approach enables users to design a wide range of scenarios.

Once the user finalizes the scenario description, \toolname{} formulates it into a structured prompt for an LLM that generates the detailed descriptions of the environment, interactive entities (e.g., victims), non-interactive elements (e.g., walls), and environmental attributes. The environment and its entities are serialized and saved into a binary file using the Pickle module \footnote{\url{https://docs.python.org/3/library/pickle.html}}, which \toolname{} uses to initialize and manage the simulation.

\begin{figure*}[!ht]
  \centering
  \includegraphics[width=\textwidth]{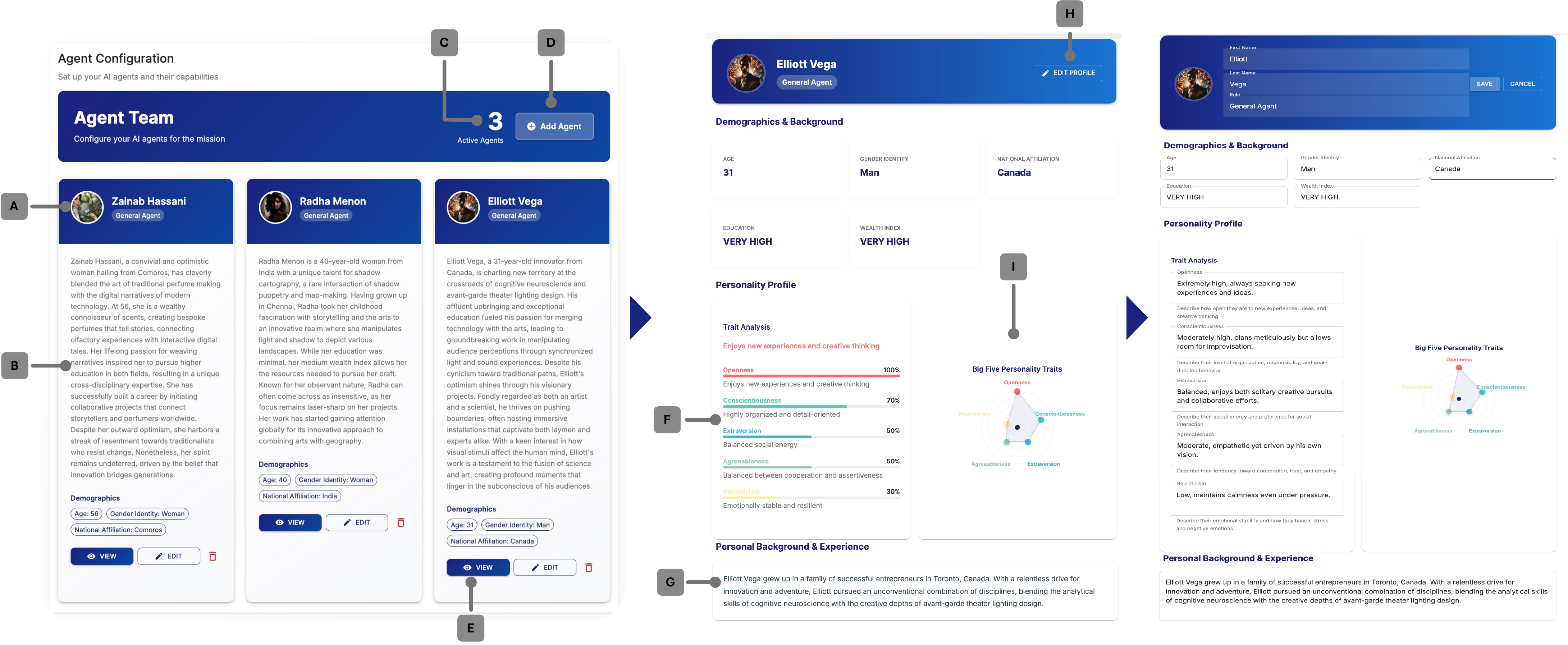}
    \caption{The agent configuration interface of \toolname{}. The system allows users to customize agents by defining demographic attributes, backgrounds, and personality traits.
    \textbf{(A)} Agent cards present each agent with the agent's name and role in the simulation.
    \textbf{(B)} Agent background and experience are presented as narrative elements that shape the agent’s behavior.
    \textbf{(C)} Displays the number of agents generated and currently active in the simulation.
    \textbf{(D)} The ``Add Agent'' button allows users to expand the team by creating new agents.
    \textbf{(E)} View allows users to access agent details, including agent demographics and personality.
    \textbf{(F)} Agent personality trait summarizes the agent’s Big Five personality profile,
    \textbf{(G)} The personality background and experience.
    \textbf{(H)} The ``Edit Profile'' button provides options to modify demographic information, background, and personality traits, and
    \textbf{(I)} The personality chart visualizes behavioral tendencies.}
  \label{fig:agent_config_steps}
\end{figure*}

\paragraph{Agent Creation} After finishing the scenario's description, \toolname{} will indicate the number of existing agents (Figure \ref{fig:agent_config_steps}c). To add a new agent, the user clicks ``Add Agent'' (Figure \ref{fig:agent_config_steps}d) and a pop-up window will appear to create the agent. By default, \toolname{} creates an agent based on the scenario's description. The users can customize the agents' names, demographics, personality traits, psychological values, behavioral characteristics, roles, and backstory memories by clicking the button ``Edit Profile'' (Figure \ref{fig:agent_config_steps}h). The interface provides text boxes to specify these characteristics (Figure \ref{fig:agent_config_steps}). Once the agent is created, the system will provide metrics to describe its personality profile. We chose the Big Five personality trait scale, displayed with explanations (Figure \ref{fig:agent_config_steps}f) and as a radar plot (Figure \ref{fig:agent_config_steps}i). Once all the details are ready, the user closes the pop-up window, and the system will display the new agent on the main dashboard. \toolname{} will display the names \ref{fig:agent_config_steps}a), AI-generated profile, and demographics \ref{fig:agent_config_steps}b). The user can view more details or edit the agent if needed \ref{fig:agent_config_steps}h). 

In the backend, \toolname{} will provide agents with contextual information of their physical environment and recent events through a sequence of conversations. The personality profiles and memory structures of the agents are encoded in embeddings and stored in a FAISS database \cite{douze2024faiss}. To align agent behavior with predefined personalities, \toolname{} uses ``Retrieval-Augmented Generation'' (RAG) to employ relevant context, such as short-term memories of current simulation and traits, through vector similarity searches on FAISS-stored embeddings \cite{lewis2020retrieval}. 

\begin{figure*}[!ht]
  \centering
  \includegraphics[width=0.7\textwidth]{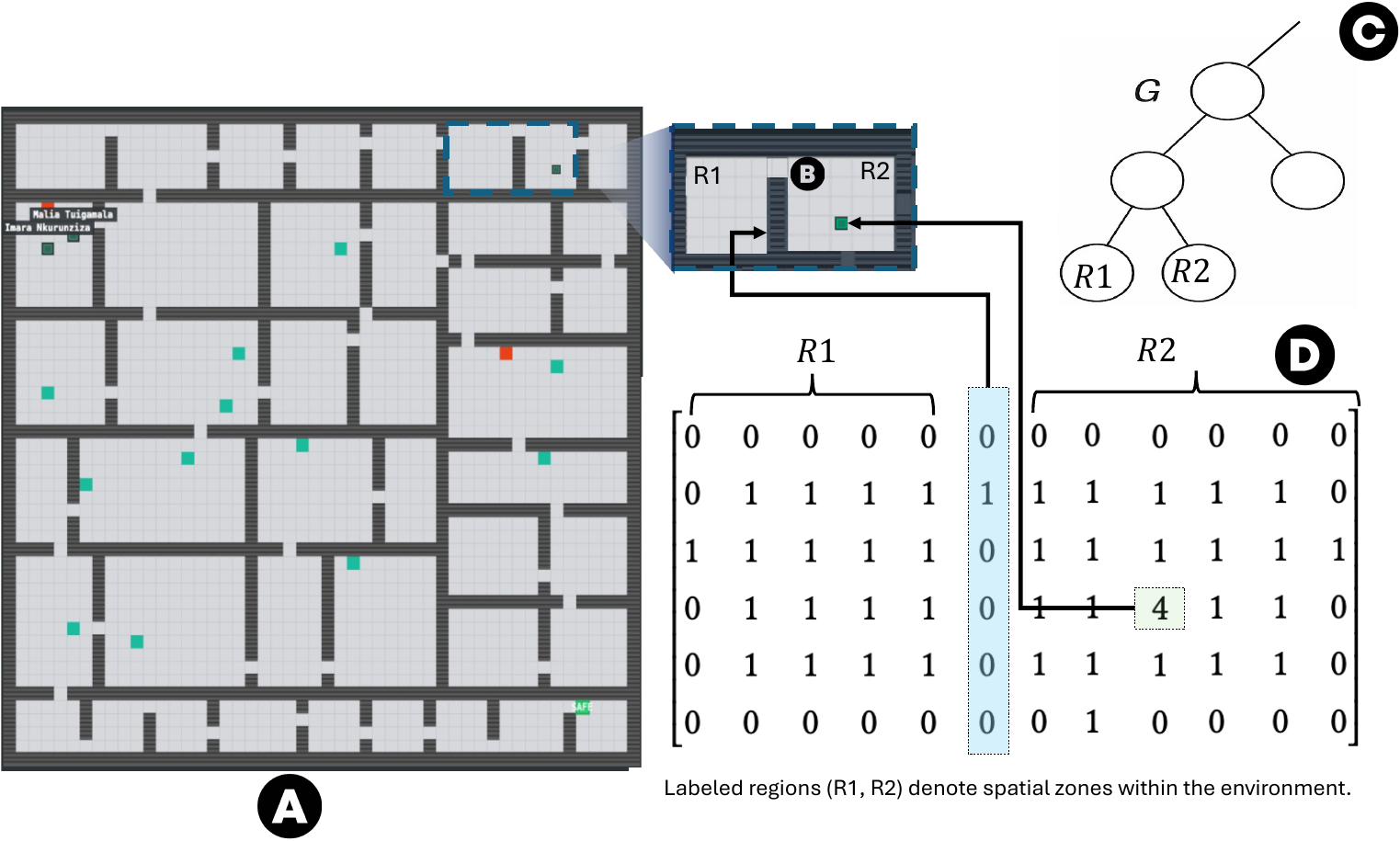}
    \caption{
    Overview of the simulation environment created in \toolname{}.  
    \textbf{(A)} The map displays the environment's physical constraints (e.g., walls), interactive entities (e.g., victims), and non-interactive entities (e.g., trees).
    \textbf{(B)} Two regions of the map environment (R1 and R2). Agents navigate the environment based on spatial constraints.
    \textbf{(C)} \toolname{} uses a partitioning algorithm to recursively subdivide the regions into a graph $\mathcal{G}$. The leaf nodes of the graph represent navigable sub-regions.  
    \textbf{(D)} The map is encoded as a binary matrix $\mathcal{M}$ that agents use to perceive their surroundings. Each environment's matrix $\mathcal{M}$ is dynamically updated throughout the simulation to reflect environmental changes and agent actions. Number `0' represents limits for the agents' movements (e.g., walls) while each agent's ID is located in the matrix. In this case, the agent with ID 4 is added to the matrix representation.
    }
  \label{fig:env_representation}
\end{figure*}

\paragraph{Environment Representation}
\label{env rep}
\toolname{} represents the environment using 2D structures (Figure~\ref{fig:env_representation}a). The system generates the environment based on spatial metrics (i.e., width, length) and regions (i.e., rooms or areas). Each region includes the boundaries (e.g., walls) and entities that the agents need to consider when navigating (Figure~\ref{fig:env_representation}b). \toolname{} maps the 2D layout to a matrix $\mathcal{M}$, which encodes the traversability within regions. In this matrix representation, \toolname{} places traversable paths between walls, representing connections between spatial regions (Figure~\ref{fig:env_representation}c). This allows agents to determine specific paths for navigation between regions while enforcing the physical constraints of the environment. To determine the connectivity and semantic relationships between rooms, \toolname{} employs a graph representation from $\mathcal{M}$. The system generates a tree graph $\mathcal{G}$ in which leaf nodes \(\mathcal{G} = \{ g_1, g_2, \dots, g_n \}\) correspond to a unique spatial region and their parent nodes represent a specific partition (Figure~\ref{fig:env_representation}d). Each node contains information about the specific region (e.g., name, characteristics), which is generated by the system. \toolname{} implements the graph $\mathcal{G}$ as an adjacency list of traversable spaces $\mathcal{C}$, representing connected rooms for the agents to transit.

\paragraph{Entities}
Lastly, \toolname{} identifies and places entities in the environment based on the scenario's description. They might be located in a specific area if the scenario provides the information, or be randomly located. Within each spatial region from $\mathcal{G}$, \toolname{} also locates entities based on the region's descriptions. \toolname{} categorizes entities into two main types: (a) \textit{Interactive Entities}: Agents can interact with these entities to perform tasks during the simulation. \toolname{} asks the agents whether they want to make an action with the entity, and it will persist the changes on the environment; and (b) \textit{Non-Interactive Entities}: These are entities that agents cannot manipulate, which provide contextual cues, and help agents navigate and understand the environment (e.g., walls).

\subsubsection{Simulation Execution Stage}
\begin{figure*}[!ht]
  \centering
  \includegraphics[width=0.8\textwidth] {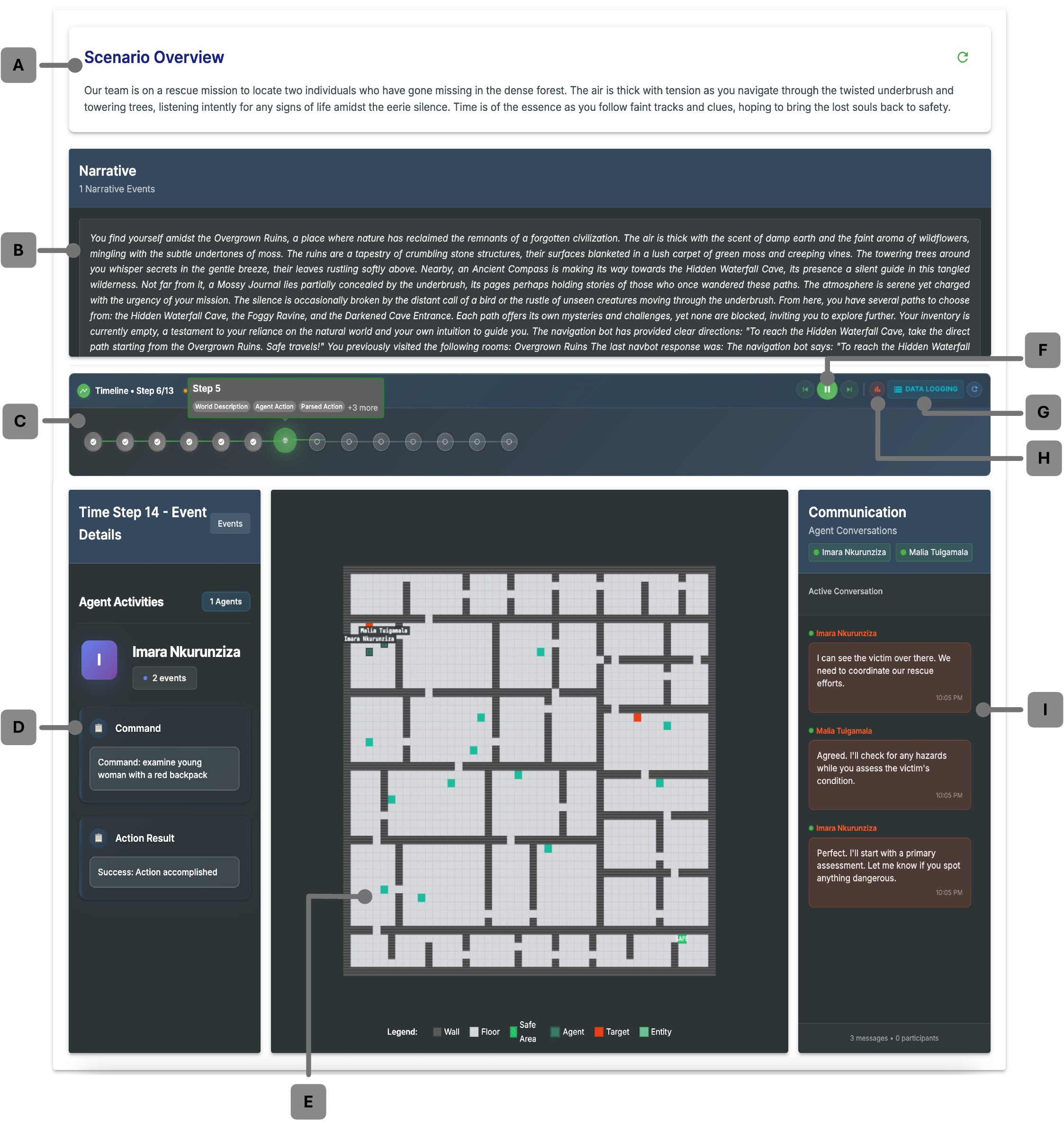}
    \caption{
    Simulation execution interface of \toolname{}. The system integrates different panels to display agent behavior and team dynamics. 
    \textbf{(A)} The scenario overview provides the mission context initially defined by the user. 
    \textbf{(B)} Narrative panel presents descriptive text updates for each active agent in the simulation, generated by the system to reflect the current state of the environment. 
    \textbf{(C)} Timeline tracks events across simulation steps; at each timestep, users can review scenario updates, environmental changes, and agent actions. 
    \textbf{(D)} Event details show agents' activity, commands, and outcomes. 
    \textbf{(E)} A 2D grid map visualizes the environment layout and agent events as they unfold during each timestep. 
    \textbf{(F)} Simulation controls allow users to pause and revisit actions. 
    \textbf{(G)} The data logging button enables users to download structured logs of the simulation in \texttt{JSON} format. 
    \textbf{(H)} Simulation analysis provides the users with qualitative insights into the simulation. 
    \textbf{(I)} The communication panel records conversations among agents. These components provide users with a comprehensive view of the simulation state, actions, and interactions.
}
  \label{fig:main_interface_sim}
\end{figure*}

Once the scenario is initialized and the agents are configured, \toolname{} proceeds to the simulation execution stage. During this stage, the system's engine manages the simulation timeline by prompting agents to act, communicate, and adapt based on their environment and goals (Figure~\ref{fig:main_interface_sim}). Agents operate autonomously, interacting through events, either actions or communications, that evolve over discrete timesteps. An event scheduling manager coordinates these interactions to ensure coherent and synchronized team behavior throughout the simulation. We explain these components in the following subsections.

\paragraph{Agent-Engine Interface}
\toolname{}'s agent-engine coordinates interactions between the simulation engine and the LLM-based agents. At each step, the system sends structured messages to every agent detailing their current location, the state of the environment, where they have been, and the outcomes from their previous actions. Using this information, agents can reason about their surroundings and decide on their next steps. The interface then prompts each agent to select among possible behaviors: executing an action, initiating communication with another agent, or remaining idle. When the agent chooses to act or communicate, \toolname{} schedules an \textit{event} based on the agent's rationale and response. Each event specifies the action type (e.g., move, pick), its duration in simulation steps, relevant contextual data, and the agents involved with the event. These interactions are implemented as API functions, allowing \toolname{} to remain flexible and integrate with different agent implementations.

\paragraph{Events}
\toolname{} models the LLM agents' actions as discrete events that occur during the simulation. The system parses the agent's response and interprets what event the agent is generating. \toolname{} supports two types of events: \textit{action} events and \textit{communication} events. Action events directly affect the state of the environment, such as moving an agent to a new location or manipulating an entity. They can take a variable amount of time to complete, depending on the type of action and agent attributes.

Communication events involve information exchanges between two or more agents. \toolname{} enables multi-turn interactions and requests the initiator agent to specify which agents both should participate in the conversation. Once the conversation starts, \toolname{} determines which agent should speak by predicting the most likely next speaker based on the ongoing conversation \cite{wu2023autogen}. It also terminates the conversation if \toolname{} believes that the information exchange becomes redundant. Agents have the choice to listen to the message or ignore it and continue with their current events.

\paragraph{Event Scheduling Manager} 
\label{Event Scheduling Manager}
To manage the multiple events' actions and durations, \toolname{} employs an event scheduling manager that executes agents' events in the correct sequence. Events can occur in parallel by advancing agents through incremental timesteps of their planned events, enabling agents to pursue separate goals simultaneously. Each event ($E_i$) is represented in event scheduling manager as a tuple \(E_i = (p_i, e_i, f_i, d_i, L_i)\) where: 

\begin{itemize}
    \item \textit{Event Type ($p_i$)}: Specifies whether the event is an action or a communication event.
    \item \textit{Execution Time ($e_i$)}: The simulation time at which the event is scheduled to occur, computed by adding an event-specific duration to the current timestep.
    \item \textit{Event Handlers ($f_i$)}: A callable, pre-defined function responsible for executing the agent’s action and applying its effects to the environment.
    \item \textit{Contextual Data ($d_i$)}: All information required to execute the event, such as the current environment state, the agent’s location in the graph $\mathcal{G}$, and relevant entities.
    \item \textit{Involved Agents ($L_i$)}: If more than one agent is involved in the event, $L_i$ represents the set of agents associated with the event. 
\end{itemize}

Algorithm \ref{alg:event_processor} illustrates the procedure the system uses to handle events. \toolname{} manages events by interactively processing agent decisions and updating the simulation state. At each time step \(t_i\), \toolname{} retrieves the next scheduled event ($E_i$) from the queue, executes the agent's action, and updates the environment based on the agent's resolution. It records state changes, notifies other agents of the outcome, and allows them to update their knowledge and decide on future actions. Each event is validated by \toolname{} against the environment state. This validation is done by asking an LLM to judge whether the agents' proposed action is possible based on the scenario's description and given the current environment's state. Invalid events lead \toolname{} to re-prompt the agent, and valid events are sent to the event scheduling component. This design also minimizes the risks of hallucinations. 

\begin{algorithm}[h]

\caption{Event Schedule Algorithm}
\label{alg:event_processor}
\KwData{Event queue $\mathcal{Q} \gets \emptyset$}
\KwResult{Process all events in $\mathcal{Q}$ in order of their time $t_i$}
\textbf{Event Definition:}
Each event $E_i$ is defined as a tuple $E_i = (t_i, f_i, d_i, L_i)$.
\BlankLine
\textbf{Add Event to the Queue:}
\SetKwProg{Fn}{Procedure}{}{}
\Fn{\textsc{AddEvent}($t_i, f_i, d_i, L_i$)}{
    Create event $E_i \gets (t_i, f_i, d_i, L_i)$\;
    Insert $E_i$ into $\mathcal{Q}$ maintaining order by $t_i$\;
    \If{$L_i \neq \emptyset$}{
        \ForEach{agent $A_j$ in $L_i$}{
            Create associated event $E_j \gets (t_i, f_i, d_i, \emptyset)$ for agent $A_j$\;
            Insert $E_j$ into $\mathcal{Q}$ maintaining order by $t_i$\;
        }
    }
}

\BlankLine
\textbf{Process Events at Time $t$:}

\Fn{\textsc{ProcessEvents}($t$)}{
    \While{$\mathcal{Q} \neq \emptyset$ \textbf{and} $\min_{E_i \in \mathcal{Q}} t_i \leq t$}{
        Extract event $E_i$ with the earliest scheduled execution time \(t_i\), such that \(t_i\) $\leq$ \(t\), is extracted from \(\mathcal{Q}\);
        
        Execute action $f_i(d_i)$\;
    }
}
\end{algorithm}

\paragraph{Navigation} Lastly, to support agents' movement and exploration in the environment, \toolname{} provides navigational cues that indicate possible routes from their current location to their target region. These cues are designed to enhance agents' spatial awareness, similar to how humans recall and use landmarks when navigating. Agents retain autonomy to incorporate this guidance into their reasoning, drawing on memory and contextual awareness to choose whether to follow the suggested route or take an alternative path. These navigation instructions are embedded in the prompts sent through the system's agent-engine interface. 

\subsubsection{Termination Stage}
The simulation concludes either when \toolname{} detects that the simulation's goal has been achieved or when the predefined time limit expires. At that point---whether successful or not---the system provides final performance metrics and allows the user to survey the agents for insights into their reasoning and teamwork dynamics. We describe the final interface and components to assist users in analyzing the final simulation outcomes.

\begin{figure*}[!ht]
  \centering
  \includegraphics[width=0.7\textwidth]{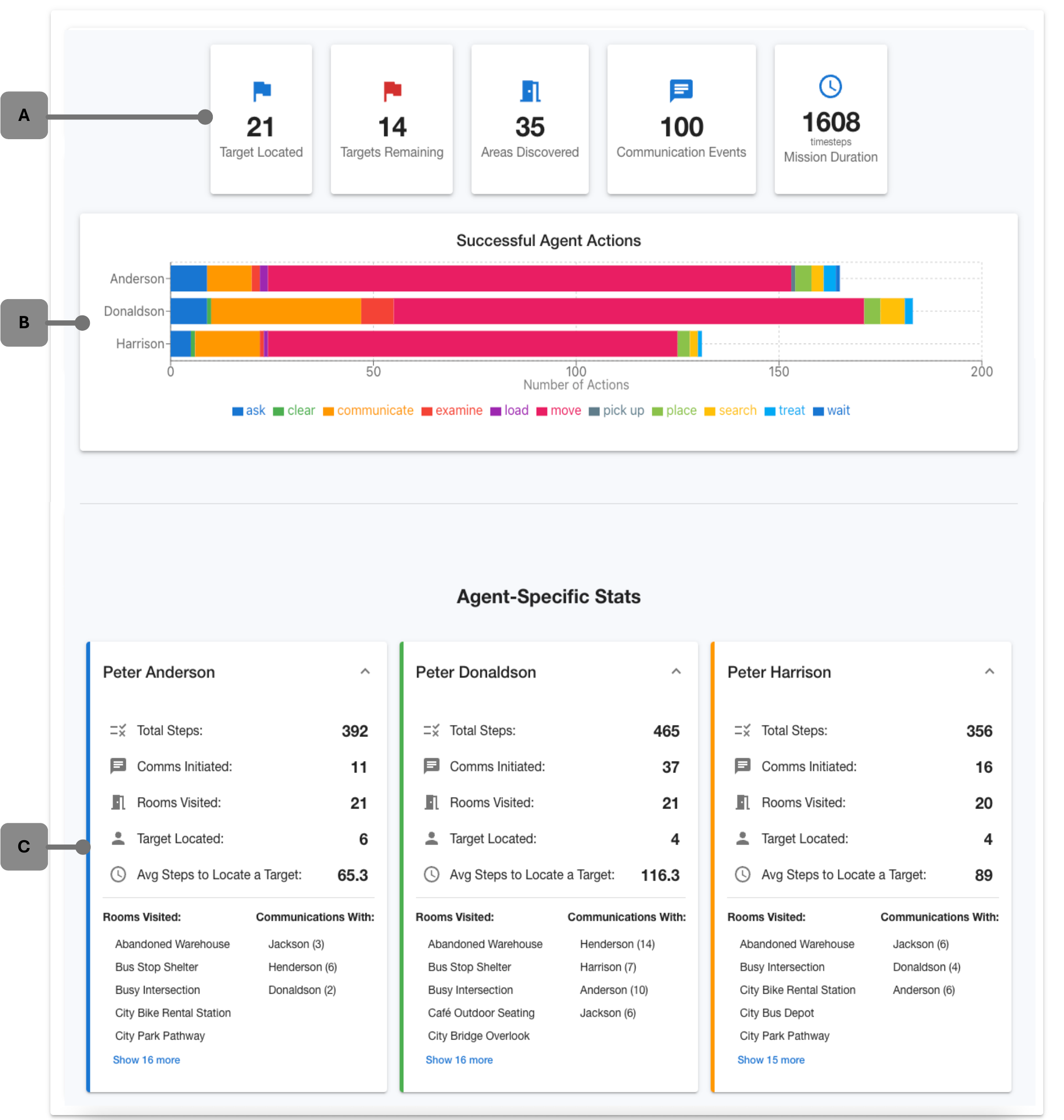}
    \caption{The \toolname{} system simulation result interface. (A) High-level mission statistics, including the number of targets located, targets remaining, areas discovered, communication events, and mission duration. (B) Visualization of agent actions categorized by type (e.g., communication, movement, pickup, search). (C) Agent-specific statistics, showing individual performance metrics such as total steps taken, communications initiated, rooms visited, and targets located. }
  \label{fig:VirT-Lab System result}
\end{figure*}

\paragraph{Results Interface}
Throughout the simulation run, the system keeps track of how many of each action are taken by each of the agents. This data is recorded and can be visualized as a bar chart in the results interface (Figure \ref{fig:VirT-Lab System result}. This screen also reports data on generic actions that agents will take regardless of the simulation, such as movement and communications initiated. Furthermore, there is data on the contribution of the agent towards the success of the mission, detailing how much exploration the agent did (area visited) and targets located.

\paragraph{Post-hoc Interviews with Agents} 
\toolname{} includes a post-hoc interview interface that enables users to directly chat with the agents after a simulation. This approach builds on methods from developmental and cognitive psychology that examine \textit{theory of mind} (ToM) capabilities, which is the ability to infer, understand, and predict the mental states of others relative to oneself \cite{Wellman2018}. Recent studies have used ToM-inspired tasks to evaluate LLMs’ capacity for logical inference and social reasoning \cite{Kosinski2024}. Extending this line of work, \toolname{} provides structured and customizable interview tools to probe ToM-like reasoning relevant to simulated team contexts.

\begin{figure*}[!htb]
  \centering
  \includegraphics[width=0.7\textwidth]{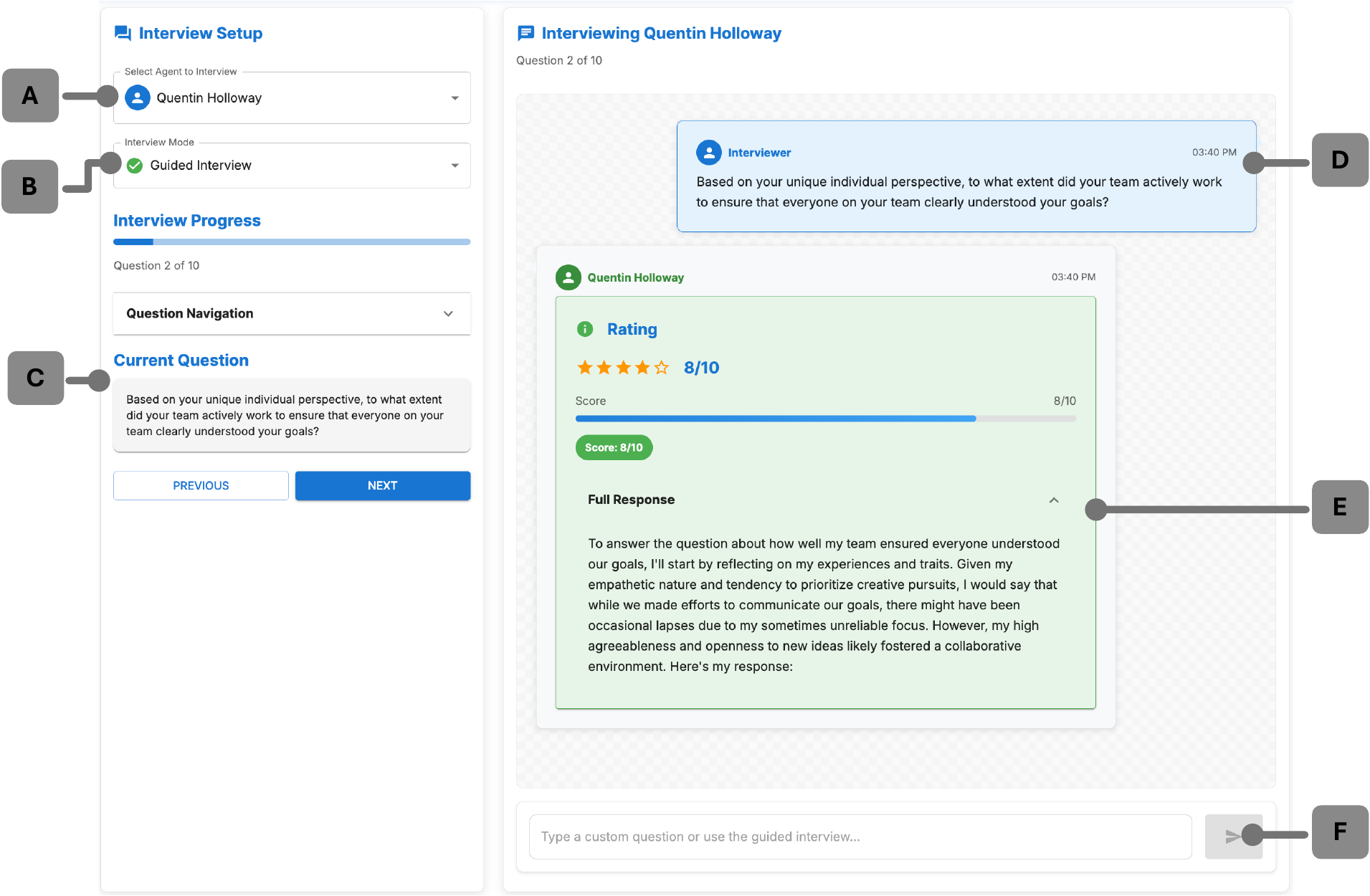}
    \caption{User interface of the guided interview panel. 
    \textbf{(A)} Agent selection for the interview, 
    \textbf{(B)} Selection of interview mode, 
    \textbf{(C)} Display of the current interview question with navigation controls, 
    \textbf{(D)} The interviewer's prompt presented to the agent, 
    \textbf{(E)} Agent’s response with rating and explanation, and
    \textbf{(F)} Input field for entering custom questions.}
  \label{fig:agent_interview}
\end{figure*}

The user selects which agent to interview (Figure~\ref{fig:agent_interview}a) and selects the type of interview (Figure~\ref{fig:agent_interview}b). The system provides a Likert-scale–based questionnaire to capture agents’ perspectives on team performance, decision-making, and mission outcomes during the simulation (Figure~\ref{fig:agent_interview}c). Each question is added to the chat with the agent (Figure~\ref{fig:agent_interview}d), with the agent responding and rating their experience (Figure~\ref{fig:agent_interview}e). Users can also ask personalized, open-ended questions to initiate a conversation with the agent (Figure~\ref{fig:agent_interview}f). The interview interface supports both guided and custom questioning to assess the agents' ability to conceptualize the knowledge, intentions, and viewpoints of other team members, offering insights into how the team is functioning. 

\begin{figure*}[!htb]
  \centering
  \includegraphics[width=0.8\textwidth]{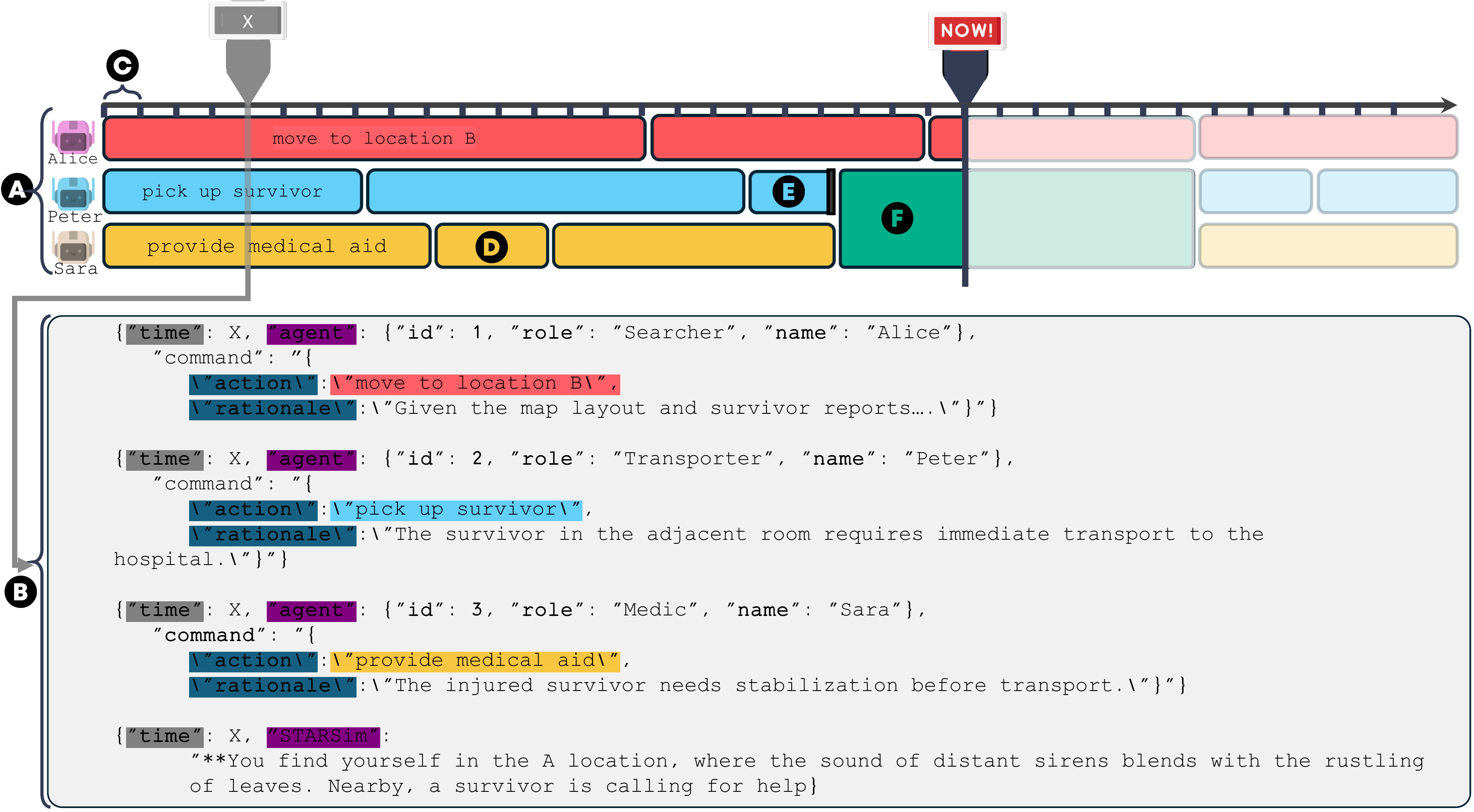}

    \caption{Visualization of \toolname{}'s event logging at a given time step. 
    \textbf{(A)} Icons representing agents in the simulation. 
    \textbf{(B)} Example JSON log entries showing how \toolname{} records each event, including the time step, agent metadata, and the agent’s command with both action and rationale.
    \textbf{(C)} A single time step within the sequential timeline of actions. 
    \textbf{(D)} Distinct colored blocks indicating different events performed by each agent, each with varying durations to reflect interactions with the environment. 
    \textbf{(E)} shows an example of an agent autonomously adapting its behavior by terminating its current event and joining another agent's event (e.g., communication).\textbf{(F)} The active task (e.g., communication) involving collaboration between two agents. }

  \label{fig:event-logging}
\end{figure*}

\paragraph{Simulation Data Logging}
Lastly, since users may have diverse research goals and require flexibility in evaluating the final simulation, \toolname{} complements its built-in analysis with structured logging in a JSON format. At every timestep, \toolname{} records the full state of the environment using a JSON array format (Figure~\ref{fig:event-logging}g), enabling users to conduct personalized analyses of the team's performance, agent-environment interactions, and team dynamics.

\subsubsection{Implementation}
We implemented the \toolname{} backend using Python and OpenAI GPT-4o-mini. We employed the \texttt{Comp-HuSim} framework to create the agents \cite{fan2024comp}. We generated the 2D environment using a binary partitioning algorithm. The environment and entities were serialized and saved using the Pickle module\footnote{\url{https://docs.python.org/3.13/library/pickle.html}}. This allowed us to store complex objects, such as the state of the environment, as binary data. We modeled the event scheduling system as a priority queue, which was implemented in Python as a heap\footnote{\url{https://docs.python.org/3.13/library/heapq.html}}. For the web interface, we used React\footnote{\url{https://react.dev/}} to enable visualization and user interaction. The environment's map is rendered using Phaser\footnote{\url{https://docs.phaser.io/phaser/getting-started/what-is-phaser}}, a game development library that supports 2D map representation and real-time updates. We used HTTP REST API\footnote{\url{https://developer.mozilla.org/en-US/docs/Glossary/REST}} to facilitate asynchronous updates between the simulation backend and the user interface.

\section{Evaluation} 
\label{Evaluation}
To evaluate the flexibility and utility of \toolname{}, we conducted four complementary evaluations: a ground truth study, a scalability analysis, an ablation study, and a user study. Together, these evaluations aim to demonstrate the robustness of the system for simulating human teams' behavior and highlight its value as a research tool. In particular, our goal is to study how \toolname{} can be used by researchers with varying levels of expertise to investigate social science hypotheses and gain deeper insight into team dynamics under various conditions.

\subsection{Ground Truth Evaluation}
\label{Simulated Scenario from ASIST dataset}
We used the Artificial Social Intelligence for Successful Teams (ASIST) dataset ~\cite{ASU/QDQ4MH_2022}, collected from a human-subject experiment study designed to evaluate team dynamics. The ASIST dataset is suitable for this evaluation because it offers rich data capturing real human team behaviors and interactions in a controlled collaborative environment. This dataset is also open-access and free. The study involved human teams of three members collaborating on a simulated urban search and rescue task on Minecraft \cite{amresh2023minecraft}, where agents provided advice to improve the state of the team (e.g., motivation), processes (e.g., coordination), and results of the mission (e.g., score of the game). The behavioral logs, survey responses, communication transcriptions, and environmental states provide a detailed baseline against which we use to assess the accuracy and realism of \toolname{}’s simulated team dynamics.

We adopted one scenario from the dataset, in which the team's primary goal is to locate and rescue 35 victims scattered throughout the environment. The scenario involves three distinctive roles for the human participants: \textit{Transporter}, \textit{Medic}, and \textit{Engineer}. The Transporter's mission is to move the victims from their starting location to the hospital in the most rapid manner. The Medic is required to assess and stabilize any victim who was designated as critical prior to transportation. Lastly, the Engineer is in charge of removing obstacles blocking paths to new locations. The rescue team explores the environment by moving between rooms and keeping track of the locations they have visited. Agents start in a room labeled as the `Hospital'. The team members are initially unaware of the victims' locations. Agents coordinate to find the nearest victims and explore the space. Agents must bring the victim back to the `Hospital' room, and each agent may transport only one victim at a time. The simulation ends when all victims are rescued or the maximum limit of simulated time steps has been reached (i.e., 2,000).

After each run of the simulated experiment, \toolname{} gives each agent a post-simulation ego survey similar to the survey given to human participants in the original ASIST study (see appendix \ref{Post-hoc agent interview}). The original survey was developed for the ASIST study, and in our work, we adapted and modified it to fit our study. The modified post-simulation ego survey uses a 10-Likert-scale survey, and includes questions about each team member's self-assessment of their own performance as well as their team's performance within the simulation. The survey assesses six key team functioning dimensions commonly studied in Human-AI teaming: team communication, team coordination, trust in team, emerging leadership, collective self-efficacy, and team processes \cite{vats2025humanai}. These dimensions capture how team members interact, align goals, and adapt to dynamic challenges.

To assess how capable \toolname{} was in approximating the ground truth data, we ran 20 simulations based on the specifications of the real participants performing their specific roles. We employed average scores for participants' traits and outcomes to model the simulated agents. To introduce variability in the simulations, we ran ten simulations with agents having high trust levels with their teammates and another ten simulations with agents having low levels of trust, as trust has been shown to significantly impact team coordination \cite{parasuraman1997humans, choung2023trust, schelble2025examining, schelble2023effect}.

\subsection{Scalability Evaluation}
To evaluate the scalability of \toolname{}, we conducted systematic stress tests under varying simulation conditions. Specifically, we examined system performance when increasing the number of agents per team and the size of the environment (i.e., environmental complexity), using the scenario described in Section~\ref{Simulated Scenario from ASIST dataset}. In this evaluation, all agents were assigned a Transporter role, and all victims were in stable condition. All experiments were carried out on a Linux server (Ubuntu 22.04) equipped with 125 GB RAM and an Intel(R) Core(TM) i7-7820X CPU. Below, we describe in detail the experimental variations used to assess scalability.

\subsubsection{Team Size}
We evaluated system scalability with respect to team size by increasing the number of active agents per simulation, ranging from two to five agents per team. Each additional agent introduced more inter-agent interactions and decision-making processes, further contributing to system load and complexity. Each test configuration was executed three times to account for variability introduced by agent decision-making.

\subsubsection{Environment Size}
We designed three environments of increasing difficulty by scaling the grid size to $30\times30$ (Low complexity), $40\times40$ (Medium complexity), and $50\times50$ (High complexity), and by randomizing the placement of 35 victims. The maps' layouts and room configurations were the same in their respective configurations. As the environment size increased, agents were required to traverse longer distances and navigate more complex paths to locate and rescue victims, thereby increasing the computational demands on the system (Figure~\ref{fig:base_maps_only.pdf} in the Appendix illustrates the three map configurations used in this evaluation).

\subsubsection{Measurements}

\begin{itemize}
    \item \textbf{Number of Rescued Victims:} The total number of victims successfully rescued by the team during each simulation. A victim was considered rescued if it was located by an agent and transported to the designated hospital area within the simulation time limit. 
    \item \textbf{Number of Action Events:} The total number of actions executed during the simulation, reflecting the dynamics of agent-agent and agent-environment interactions. 
    \item \textbf{Number of Communication Events:} The total number of communication exchanges between agents during the simulation, including both dyadic and multi-agent conversations.
    \item \textbf{Simulation duration:} Total time step required to complete each simulation, measured from the start of agent deployment to the completion of the mission (all victims rescued) or the expiration of the simulation time step. To ensure consistency, the duration of the simulation was averaged across all runs for each configuration.
\end{itemize}

\subsection{Ablation Study}
To evaluate the contribution of \toolname{}'s core architectural components, we conducted an ablation study in which we systematically disabled components and compared agents' performance against the full system. Specifically, we focus on four core components: 

\begin{itemize}

\item \textbf{No Navigation Assistance:} Agents were restricted to only seeing immediately adjacent rooms, without access to efficient path planning across the environment. This condition isolates the impact of navigation efficiency on overall team performance. 

\item \textbf{No Communication:} Agents were restricted from exchanging messages, including both dyadic and multi-party conversations. This condition isolates the role of communication in team coordination.

\item \textbf{No Event Scheduling Manager:} The event scheduling system was disabled, causing agent events to be executed in a round-robin approach. This condition evaluates the role of temporal coordination in executing events in parallel and its effect on producing coherent team dynamics.

\item \textbf{No Agent Memory:} Agents operated without access to their memories stored in the FAISS database. We disabled the agents’ ability to create or update memories during the simulation and turned off context retrieval through RAG. This condition allows us to evaluate how short-term and long-term memory impacts context-aware decision-making and team collaboration.

\end{itemize}

Based on these combinations, we evaluated \toolname{}’s performance using mission success rate as the evaluation metric. Each ablation configuration was run on the easy map condition and was conducted three times.

\subsection{User Study}
We conducted a user study to evaluate the usability, interpretability, and perceived realism of the \toolname{} system. The goal of this study was not to measure the system performance, but rather to prompt likely users to reflect on the ease of use and the potential of LLM-based team simulations for research. In particular, we aimed to capture participants' interpretations, expectations, and perceived affordances of \toolname{} as a research tool. The study was reviewed and approved by \anon{the University of Notre Dame}'s IRB, and all participants signed an online consent form on Qualtrics prior to participation. 

The study addressed three primary research questions:
\begin{itemize}
\item[\textbf{RQ1:}] Can users use \toolname{} effectively and intuitively to create and configure team simulations as intended?
\item[\textbf{RQ2:}] What benefits and limitations do users perceive when simulating human team dynamics using \toolname{}?
\item[\textbf{RQ3:}] How do users’ levels of simulation expertise influence their experiences when using \toolname{}?
\end{itemize}

\subsubsection{Participants}
We recruited 12 participants from simulation and team science communities through direct invitations, academic mailing lists, and social media posts. This sample size is supported by previous usability research \cite{nielsen1993mathematical}. Our final sample consisted of participants from STEM-related fields, including researchers, graduate students, undergraduate students, and professionals with varying levels of simulation expertise. Based on self-reported experience with agent-based modeling and LLM-based agents, three participants were classified as experts, four participants as intermediate, and five participants as novices with little or no prior experience in these areas. To ensure accurate classification, two members of the research team reviewed the interview responses of the participants describing their previous experience with simulation systems. Demographic information and experience levels are summarized in Table~\ref{table:participant_info}.

\subsubsection{Procedure}
Each session lasted 60 minutes and was conducted in person in a university research lab or virtually through Zoom, depending on participant availability. We compensated participants with a \$20 digital gift card for their time. Our study was conducted in three phases:

\begin{enumerate}
\item \textbf{Pilot Study} We began by conducting a pilot study with three recruited participants. These participants had prior experience with agent-based systems or team science and were invited to participate in an early version of the \toolname{} system. During these sessions, they explored the system’s features, ran multiple scenarios to simulate team dynamics, discussed its potential to emulate team dynamics, and provided feedback on usability issues and interface design limitations.

\item \textbf{System Refinement} Based on the pilot study, we added brief tool tips and minor text edits. These adjustments did not significantly change the interface, user workflow, simulation timing, or study procedures. 

\item \textbf{Formal Study}: We conducted structured usability sessions with nine additional participants. Before each session, participants received a brief introduction and completed a background questionnaire assessing their experience with team simulations, programming, and AI systems.

\end{enumerate}

\subsubsection{Post-study Survey}
After using \toolname{}, participants evaluated the usability of the system using the SUS scale \cite{brooke1996sus} (Cronbach’s $\alpha$ = 0.75). Trust in the system was assessed using a 5-point Likert scale adapted from \cite{hoffman2018metrics} ($\alpha$ = 0.83). Explanation satisfaction was measured using items adapted from \cite{hoffman2018metrics} that evaluated whether the explanations provided by the system were clear and supported the participants’ understanding of the simulation results ($\alpha$ = 0.86). 

The realism of the system and environment was evaluated with items adapted from Witmer and Singer's presence questionnaire \cite{witmer1998measuring}, which captured participants’ perceptions of the simulated environment ($\alpha$ = 0.71). Perceived adoption was measured with items adapted from \cite{venkatesh2008technology} to evaluate participants’ willingness to use the system in future research or professional contexts ($\alpha$ = 0.85). Additionally, we created a 5-point Likert scale with two items to evaluate the realism of simulated agent and team behaviors. The items included items such as \textit{``The agents’ behaviors were believable and consistent with their defined roles''} and \textit{``The agents’ communication resembled interactions in real teams''} ($\alpha$ = 0.90). 

Finally, we included open-ended questions to examine the perceived benefits and limitations of \toolname{}. The questionnaire responses for each scale were averaged into a single score per participant's expertise level.

\subsubsection{Post-study Interviews}
Each session followed a semi-structured format. Participants were asked to come up with short team simulation scenarios (e.g., \textit{``A team of first responders battles the roaring wild; With the smoke thick in the air, they coordinate efforts to evacuate three residents.''}) The participants then ran their scenarios using \toolname{}. In four cases, the scenarios were too complex to complete within the study session, so participants were encouraged to adjust their scenarios to reduce complexity. While using the system, participants were asked to think aloud. After the simulation phase, participants completed a post-study questionnaire and a formal interview (see Appendix \ref{Appendix: Post-Study Questionnaire}). During interviews, we follow our study protocol with follow-up questions when needed. Specifically: 

\begin{enumerate}
    \item Baseline questions (e.g., how participants’ perceptions of the system and its capabilities changed before and after using the system).  
    \item Usability and workflow experiences questions, focusing on areas where the system felt clear, confusing, or difficult to control.  
    \item Questions about their learning curve and the perceived realism of the system in emulating human team behaviors, including how participants interpreted agents’ interactions and decisions.  
    \item Comparison questions between \toolname{} and other systems they had used, including how the system fit with their existing tools and what might encourage long-term use. 
    \item If time permitted, we further asked open-ended reflections to capture insights that were not addressed by our specific questions.  
\end{enumerate}

\begin{table*}[h]
\centering
\caption{Participant Information}
\begin{tabular}{|l|l|p{6cm}|l|}
\hline
\textbf{ID} & \textbf{Simulation Expertise Level} & \textbf{Simulation Tools Used} & \textbf{Occupation} \\
\hline
\rowcolor{novicegreen!40}
P01 & Novice & Not provided & Academic \\

\rowcolor{novicegreen!40}
P02 & Novice & Unity & Academic \\

\rowcolor{interblue!40}
P03 & Intermediate & Custom-built simulations (Python, Java, etc.) & Academic \\

\rowcolor{expertred!40}
P04 & Expert & Custom-built simulations (Python, Java, etc.) & Professional \\

\rowcolor{interblue!40}
P05 & Intermediate & Custom-built simulations (Python, Java, etc.) & Academic \\

\rowcolor{expertred!40}
P06 & Expert & NetLogo, Unity, ML-Agents & Academic \\

\rowcolor{novicegreen!40}
P07 & Novice & Custom-built simulations (Python, Java, etc.) & Student \\

\rowcolor{expertred!40}
P08 & Expert & NetLogo, MATLAB, Simulink, Custom code (Python, Java, etc.) & Academic \\

\rowcolor{novicegreen!40}
P09 & Novice & Not provided & Student \\

\rowcolor{novicegreen!40}
P10 & Novice & Not provided & Student \\

\rowcolor{interblue!40}
P11 & Intermediate & Cadence Specter & Academic \\

\rowcolor{interblue!40}
P12 & Intermediate & Cadence Virtuso, Mentor Graphics, Cadence Specter & Academic \\
\hline

\end{tabular}
\label{table:participant_info}
\end{table*}

We recorded these interviews and transcribed them for ease in coding. Two of the researchers met to code a few transcripts to establish an initial codebook. The analysis aimed to identify recurring themes related to usability, perceived realism, and simulation control. When needed, the researchers would refer to the recordings directly to view the participants' interactions with the system. The researchers then coded the rest of the interviews independently and met again to review and establish the final codebook. Differences between the coding were resolved through discussion between the researchers until consensus was achieved. The researchers then grouped the codes according to themes \cite{braun2006using} (see Appendix \ref{tab:codebook_overview}). The quotes are color-coded according to their level of expertise: \N{novice}, \I{intermediate}, and \E{expert}.
\section{Findings}
\label{Result}

\subsection{Ground Truth Analysis}
\begin{figure*}[!htb]
    \centering
    \includegraphics[width=0.7\textwidth]{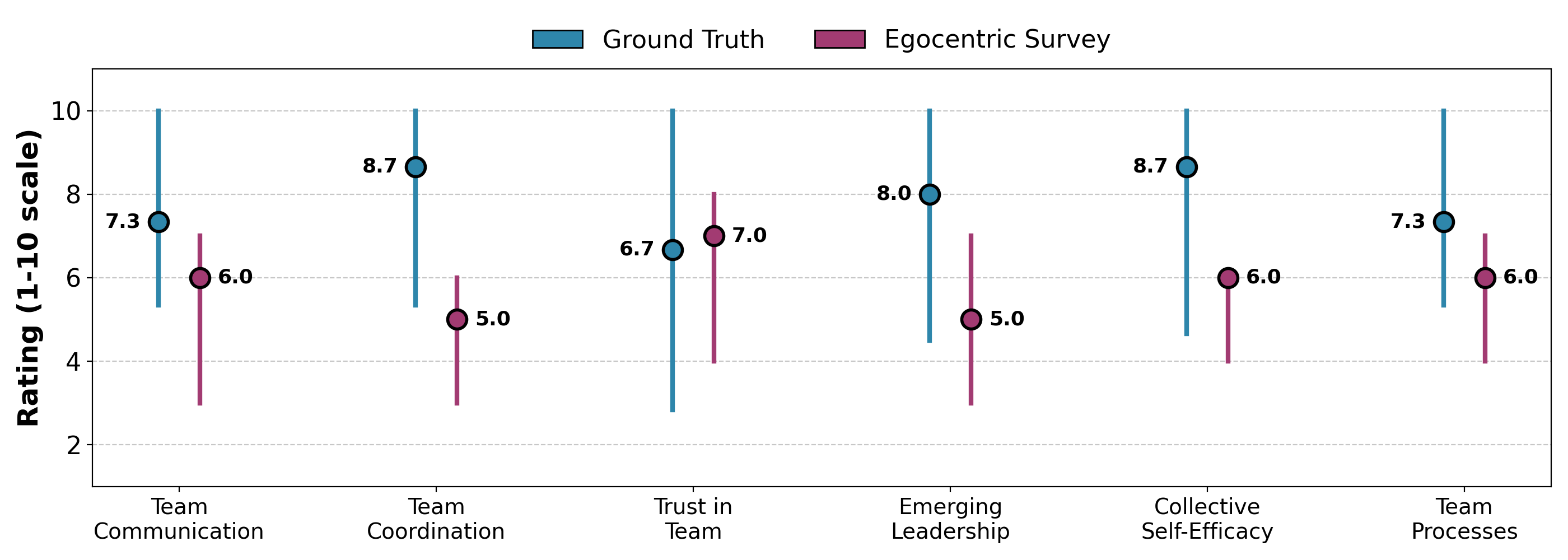}
    \caption{Average scores of real participants and simulated participants. Real participant values were originally measured on different scales and rescaled to a 1–10 scale for easy comparison. The values of the real participants were consistently higher than those of the simulated participants measured using ego surveys.}
    \label{fig:survey-data}
\end{figure*}

We found that the attributes embedded in the agents significantly influenced the team dynamics during the simulation. As shown in Figure \ref{fig:survey-data}, the ground truth teams consistently provided higher ratings across all six key team functioning dimensions. In contrast, the simulated agents consistently rated lower on these dimensions. This suggests that while the simulated teams exhibited team dynamics, their strength was lower than that of real human teams. The largest discrepancies were observed in team coordination and emerging leadership. Human participants rated their coordination and leadership dynamics higher. However, simulated agents rated them lower. Interestingly, trust in the team exhibited the smallest gap between ground truth and simulation. This finding suggests that the simulated agents were able to approximate the way human participants perceived the reliability and dependability of their teammates.

\subsection{Scalability Analysis}
Table~\ref{tab:scalability_results} shows how \toolname{} scales across different map complexities and team sizes. Overall, simulated teams successfully rescued nearly all victims in easy and medium maps, whereas the performance degraded in the hardest maps. Simulated teams with two agents rescued 20.67 victims on average, far from the total of 35. Simulation duration decreased as the number of agents increased in easy and medium maps, indicating that larger teams could complete the missions more efficiently. However, adding more agents also increased the total number of action events, though the number of actions per agent decreased, suggesting a division of labor across team members. Communication events remained relatively stable across team sizes, with only modest increases in medium and hard maps. The results highlight how \toolname{} can manage multi-agent simulations across varying levels of difficulty, while also revealing trade-offs between efficiency, workload distribution, and mission success when scaling team size and environment complexity.

\begin{table*}[!htb]
\centering
\caption{Scalability results across map complexity levels and agent counts. Values are reported as mean~$\pm$~standard deviation.}
\renewcommand{\arraystretch}{0.9}
\begin{tabular}{lccccccc}
\toprule
\makecell{\textbf{Environment}\\\textbf{ Agents}} & 
\makecell{\textbf{Victims}\\\textbf{Rescued}} & 
\makecell{\textbf{Simulation Duration}\\\textbf{(Steps)}} & 
\makecell{\textbf{Action}\\\textbf{Events}} & 
\makecell{\textbf{Action Events}\\\textbf{per Agent}} & 
\makecell{\textbf{Communication}\\\textbf{Events}} \\
\midrule
\rowcolor{gray!10}\multicolumn{6}{l}{\textbf{Easy Map Complexity}} \\
2 Agents & 35 & 1,902.67 $\pm$ 84.98 & 1,431.0 $\pm$ 109.7 & 715.5 $\pm$ 54.9 & 73.33 $\pm$ 9.03\\
3 Agents & 35 & 1,309.67 $\pm$ 94.32 & 1,426.3 $\pm$ 91.2 & 475.4 $\pm$ 30.4 & 69.67 $\pm$ 22.4\\
4 Agents & 35 & 904.33 $\pm$ 40.66 & 1,262.3 $\pm$ 115.8 & 315.6 $\pm$ 29.0 & 63.67 $\pm$ 9.1\\
5 Agents & 35 & 730.67 $\pm$ 44.79 & 1,313.3 $\pm$ 80.9 & 262.7 $\pm$ 16.2 & 61 $\pm$ 17.45\\
\midrule
\rowcolor{gray!10}\multicolumn{6}{l}{\textbf{Medium Map Complexity}} \\
2 Agents & 33.33 & >2,000 & 1,290.0 $\pm$ 12.3 & 645.0 $\pm$ 6.2 &  29.67 $\pm$ 0.94\\
3 Agents & 35 & 1,584.33 $\pm$ 51.91 & 1,842.3 $\pm$ 12.5 & 614.1 $\pm$ 4.2 &  35.33 $\pm$ 18.19\\
4 Agents & 35 & 1,184.67 $\pm$ 20.74 & 2,543.7 $\pm$ 53.5 & 635.9 $\pm$ 13.4 &  32.33 $\pm$ 2.87\\
5 Agents & 35 & 995.33 $\pm$ 20.89 & 2,665.3 $\pm$ 109.0 & 533.1 $\pm$ 21.8 &  42 $\pm$ 12.03\\
\midrule
\rowcolor{gray!10}\multicolumn{6}{l}{\textbf{Hard Map Complexity}} \\
2 Agents & 20.67 & >2,000 & 1,380.7 $\pm$ 48.8 & 690.3 $\pm$ 24.4 & 24.67 $\pm$ 10.14\\
3 Agents & 31 & >2,000 & 1,614.3 $\pm$ 35.7 & 538.1 $\pm$ 11.9 & 26.33 $\pm$ 5.31\\
4 Agents & 34 & >2,000 & 1,583.3 $\pm$ 51.8 & 395.8 $\pm$ 12.9 & 27.67 $\pm$ 4.78\\
5 Agents & 35 & 1,686.33 $\pm$ 39.97 & 1,694.0 $\pm$ 7.3 & 338.8 $\pm$ 1.5 & 31.33 $\pm$ 6.94\\
\bottomrule
\end{tabular}
\label{tab:scalability_results}
\end{table*}

\subsection{Ablation Analysis}
Our results indicate that the full architecture achieved a reliable overall performance. In this condition, agents successfully rescued all 35 victims with an average of 730.67 steps (SD = $\pm 44.79$). Interestingly, when the communication component was removed, performance did not decline; in fact, agents achieved comparable rescue success with slightly higher efficiency (718.33 steps, SD = $\pm 34.99$). This small improvement is likely due to the absence of time spent exchanging messages, suggesting that communication was not a critical factor for coordination in this relatively simple search task. By contrast, removing memory from the agents substantially reduced the teams' efficiency. Although all 35 victims were rescued, agents required 1,061.66 steps on average (SD = $\pm 42.13$), showing the critical role of memory in preventing redundant exploration. The most severe performance decline occurred when removing the event scheduling manager and the navigation assistance components, which are core system features. Without the event scheduling manager, the agents rescued only 20.67 victims on average (SD = $\pm 0.47$), requiring over 2,000 steps and covering just 61.4\% of the environment. Lastly, eliminating the navigation feature from the agents further degraded team performance: only 14 victims were rescued in each run (SD = $\pm 0.00$), covering only 38.6\% of the environment and exceeding 2,000 steps. 

\subsection{User Study}

\subsubsection{Post-study Questionnaire}
\begin{figure*}[!htb]
    \centering
    \includegraphics[width=0.8\textwidth]{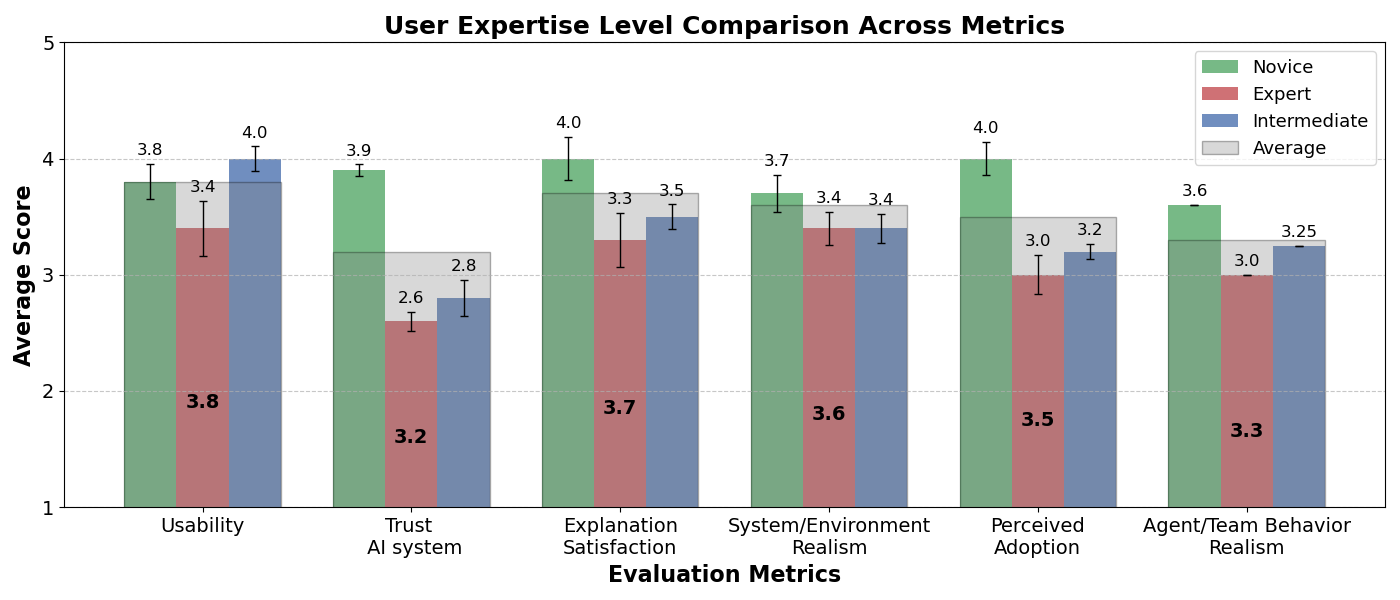}
    \caption{Participants’ survey responses. Error bars denote standard errors.}
    \label{fig:UsabilityStudyPlot}
\end{figure*}

Figure \ref{fig:UsabilityStudyPlot} summarizes the results of the post-study questionnaire. Overall, the participants rated the usability of \toolname{} positively ($M = 3.8$), although clear differences emerged by the experience level. Novice participants reported a mean usability score of 3.8 ($SD = 0.48$), intermediate users gave the highest ratings ($M=4.0, SD = 0.33$), and experts rated usability lower ($M=3.4, SD = 0.75$). Trust in the system showed similar differences. Across all participants, the mean trust score was 3.2, with novices reporting the highest trust scores ($M = 3.9, SD = 0.10$), intermediates lower trust ratings ($M = 2.8, SD = 0.31$), and the experts the lowest ($M = 2.6, SD = 0.17$). 

Satisfaction with the system's explanations was also mixed. The overall rating was 3.7, with novices reporting the highest satisfaction ($M = 4.0, SD = 0.53$), intermediates giving average ratings ($M = 3.5, SD = 0.30$), and experts reporting the lowest ($M = 3.3, SD = 0.65$). The participants' perceptions of the system and environment realism were similarly rated ($M = 3.6$) across all levels of experience. The novices reported an average of 3.7 ($SD = 0.16$), while expert users ($M = 3.4, SD = 0.14$) and intermediate users ($M = 3.4, SD = 0.12$) reported lower scores. The perceptions of system and environment realism were assessed as moderate across the various groups, with an overall mean rating of 3.6. Novices rated realism the highest ($M = 3.7, SD = 0.16$), followed by intermediates ($M = 3.4, SD = 0.12$) and experts ($M = 3.4, SD = 0.14$) who gave similar, lower ratings. 

Participants' willingness to adopt the \toolname{} system in future use followed a similar pattern. On average, perceived adoption was rated at 3.5. Novices expressed the highest adoption intentions ($M = 4.0, SD = 0.37$), while intermediates provided average ratings ($M = 3.2, SD = 0.17$), and experts reported the lowest scores ($M = 3.0, SD = 0.45$). Finally, perceptions of agent and team behavior realism varied across expertise levels. The overall average rating was 3.3. While novices rated realism the highest ($M = 3.6$), intermediates ($M = 3.25, SD = 0.00$) and experts ($M = 3.0, SD = 0.00$)  reported lower intentions.

\subsubsection{Post-study Interview}
\paragraph{KF1: \toolname{}'s design workflow was intuitive and allowed users to create simulations more naturally.}
Many users found that \toolname{}'s workflow of creating a simulation scenario and then observing it unfold reflects how they would naturally reason through a team simulation. The hierarchical setup---starting with defining an initial scenario context, then specifying each agent with detailed demographic information, personality traits, and roles in the simulation, and finally watching those agents act step by step---felt intuitive to them. As \N{P01} noted, “\textit{I do like this workflow \dots I just need to follow each step}”. Several users especially appreciated the ability to adjust agent characteristics, with \N{P02} remarking, “\textit{my favorite, or like the best [is] changing the personality [traits] of the agents}”. Some users highlighted that \toolname{} guided their simulation ideation, helping them envision team collaboration more clearly. \I{P03} stated, “\textit{It [the system] can be very useful, you know, in terms of turning my ambiguous idea into concrete actions that can be simulated and visualized”}. Others noted that the system helped to expand their initial scenario ideas, allowing them to explore and develop more complex team dynamics than they had originally considered. For example, \E{P08} explained that ``\textit{I liked the way it helps [me] to create teams or agents… and then simulate the scenario}''. In particular, participants used the scenario enhancement feature to complete their unfinished scenario description, or to save time and effort on how much detail to put into the scenario. Though the participants found the workflow to be logical and intuitive, they also noted that examples or suggestions built into the interface could help with ideating.

\paragraph{KF2: Users' prior simulation experience levels impact expectations and interactions with \toolname{} system}
Users' expertise backgrounds shaped both their expectations and their interactions with \toolname{}. Novice users often entered the study with limited expectations, but they reported that the system’s workflow offered an accessible learning curve. Several novices noted that \toolname{} exceeded their expectations once they tried it, increasing their confidence to use the system again. \N{P09} \textit{``It was my first time running a team simulation with large language model agents \dots some interface components were a little confusing, but the total workflow was good for me and helped me understand how team simulation works.''}

Users who were more familiar with simulation tools compared \toolname{} to prior systems they had used. They highlighted the ease of scenario setup but also expressed interest in having more flexibility and customization in the simulation process. \I{P03}: ``\textit{I imagine more people will learn it. I feel confident using it, though I still have things to learn. I would like more flexibility — for example, changing an agent’s personality by adjusting the radar-map scores rather than editing text; that would make control easier}''. \I{P12} suggested that the system should allow greater control, asking whether agents could be given access to the entire environment at the beginning of the simulation, \textit{``...is there any way that [I] can, show them [agents] the entire map at [the] very beginning?''} More experienced users also had higher expectations for the system's capabilities and scope. They were cautious about the system’s utility, pointing to limitations in simulation realism and control. As one participant explained, \textit{``From what I saw in my simulation \dots, there’s clear coordination between them [agent],\dots but movements seem more random — so the sky's the limit for how much you could [improve] here.''} (\E{P06})

\paragraph{KF3: \toolname{} helped users explore a range of scenarios and team behaviors, though with limitations}
Many participants reported that \toolname{} lowered the barrier to trying many scenario variants and observing different team behaviors in action. For example:
\begin{quote}
\E{P06} \textit{``No coding is great— it [the system] lets me set up variants fast and focus on the team behavior,  type it and wait 10 minutes, whatever it took to, like, fully do it. I mean, I couldn’t have coded that in 10 minutes \dots that is a huge pro. Even when I knew NetLogo pretty well, I don’t think I could have done that in 10 minutes, so I think the biggest con [sic] is that I didn’t have to do any coding. Like I said, you literally can type up a scenario, and at test it.''} 
\end{quote}

The visualization helped users iterate and compare alternatives. \I{P11} stated, \textit{``Seeing it play out on the map—almost like a video game—made it easier to explore different [simulation]''}. Also, users noticed that configured agent personality traits carried through to agent behavior, matching what they set up. For example, \I{P12} agreed, \textit{``It [actions] also reflects the personality traits... one agent gave long, agreeable answers, the other was brief—so it matched what [I] set.''}. At the same time, participants pointed to limits that impact realism and control. \I{P03} \textit{``The [agent] interview answers felt a bit ‘GPT-like’ and didn't always match what they did''}. Similarly, \N{P01} explained, ``I was expecting like something real human [would] say ..., but [it] will help a lot, I like the idea a lot to talk to them.''  

\paragraph{KF4: Lack of \toolname{}'s transparency reduced users' trust in the simulations.}
Across expertise levels, participants questioned what information was sent to the LLM, what was returned, and how these exchanges shaped the simulation outcomes, concerns that directly influenced their trust in the system. Novices and intermediates explicitly linked clarity to trust. \I{P03} noted, \textit{``if this part can be transparent, I think my trust… [goes] up''}, while also reporting confusion when they could not tell what the system was doing. These users looked for signs \textit{``what’s happening now''} (\I{P03}), expressing unease when backend processes felt delayed. As \N{P07} remarked, \textit{``I don't know what the [system] is doing. It’s kind of like an in-house system, but I think it’s pretty common''.} Experts approached trust more analytically, focusing on how data were collected and processed by the LLM and whether simulated agent behavior aligned with assigned traits. As \E{P04} reported, \textit{``I would trust the agents more if they tied their responses to actual events rather than generic personality templates''.}
\section{Discussion}
\label{Discussion}
In this study, we designed and evaluated \toolname{}, an LLM-based system that simplifies the creation and execution of human team simulations. By leveraging LLMs, the system generates simulations capable of more sophisticated behaviors, spatial awareness, and communication patterns than those produced by traditional rule-based systems. Moreover, the system lowers the technical barriers associated with traditional simulation tools, providing an accessible, web-based interface to a wide range of users. In the following sections, we discuss the implications of our findings, reflecting on how \toolname{} advances the design of human–AI systems and identifying key opportunities for future development and ethical considerations.

\subsection{Approximating vs. Emulating Human Team Dynamics}
Our ground-truth validation highlights both the potential and limitations of using LLM-based agents to model human team dynamics. \toolname{}'s agents demonstrated several realistic team behaviors, such as coordinating to accomplish shared goals, communicating intentions, and even developing a degree of trust among teammates. These findings suggest that LLM-based agents can capture important aspects of real human teams. However, we found that these behaviors, particularly coordination and emerging leadership, were noticeably weaker in LLM-based agents compared to human teams. This finding aligns with previous work that shows a gap between LLM-agent behavior and real human behavior. For example, \cite{park2024generative} found that LLM-based agents were not fully replicating human responses. In our case, \toolname{}'s agents approximated certain aspects of team dynamics, although they do not yet completely emulate the complex social and cognitive processes that emerge in real human teams. 

We identified two main factors contributing to the weaker team dynamics. First, human teams naturally develop collective memories of past successes, failures, and role expectations, which serve as the foundation for developing team dynamics for future interactions \cite{li2023theory}. In contrast, \toolname{}’s agents lack past memory and interactions beyond the current simulated scenario. This limitation prevents them from developing deeper forms of team dynamics, ultimately weakening the emergence of strong team dynamics. Second, we observed that communication decreases as the complexity of the simulated scenario increases. This pattern is consistent with findings from Grötschla et al. \cite{grotschla2025agentsnet}, who reported that LLM agents often reduce communication under scaling conditions. In our scalability evaluation, as the simulated scenario became more complex, the number of communication events dropped. With fewer communication exchanges, agents had limited opportunities to align their goals and form deeper social interaction during the simulated scenario, which restricted the development of human-like team dynamics.

Despite these limitations, \toolname{} provides a valuable complementary approach for studying team dynamics. Like many simulation tools, it is not a replacement for human-subject experiments but rather a sandbox environment where variables can be isolated and manipulated under controlled conditions. Achieving an exact emulation of a human team is beyond current AI capabilities, given the complexity and variability of human nature \cite{pan2024agentcoord, park2024generative, zhou2023sotopia}.

\subsection{Guided Workflow and Cognitive Load in Team Simulation}
The structured workflow design of \toolname{} guides users through the process of creating different parts of a team simulation, making the experience more accessible to most users. In particular, users found several aspects of the workflow helpful in understanding the scope of the system. For instance, in the scenario description part, the ability to \textit{``enhance''} their initial scenario description helped participants to improve and refine simulated scenarios. Breaking the simulation setup into several discrete guided steps was crucial in helping users make a logical sense of the requirements for creating a simulation. One participant (P03) noted that the system helped clarify their previously ambiguous thoughts, saying that before using the \toolname{}, they ``\textit{cannot imagine how they're gonna do it, this system kind of instantiates my ambiguous thoughts}''. This scaffolding was especially important for \toolname{}, as the text-based LLM system can quickly become cognitively overloading without some structure to guide users \cite{epperson2025interactive}. 

Despite these positive aspects, overall usability metrics reflect only a slightly above-average experience. Participants reported an inconsistency in the simulation loading, especially when it came to the rate of each time step generation during simulation execution. Due to the architecture of the \toolname{}'s backend, which involves sending and receiving data through API calls to the LLM, simulation updates differ by event type, so the front end presents time steps in uneven intervals. This behavior aligns with findings on LLM performance, where latency between processing the input prompt and generating tokens is affected by factors such as prompt length and network delays, creating irregular update patterns that are visible to end users \cite{yu2022orca, agrawal2024etalon}. As a result, several users found this tedious and wanted to be able to ``\textit{force it to move to the next step}'' when waiting (P02), and another described the time-step display as ``\textit{not that smooth. Sometimes it's just, like, flash[ing]}'' (P07). Participants also noted that generating the agents and advancing the simulation took longer than they expected, with no accurate indication of loading progress, as (P08) stated, ``\textit{Agent creation takes more time than expected, like, there is no feedback, like, you have to wait, I mean, how many minutes, or how many seconds}''. Overall, while the design of \toolname{} helped mitigate many inherent challenges in setting up team simulation, there were still notable usability challenges. The slow and uneven loading of the simulation leads to user frustration with the system's usability.  

\subsection{Benefits and Limits of LLM Team Simulation}
\toolname{} system enabled more natural interaction and much faster scenario prototyping, though its advantages were qualified by issues of realism, consistency, trust, and performance. A major benefit users noted was the ability to set up and explore team scenarios quickly compared to other simulation tools. Several reported that \toolname{} let them \textit{'set up variants fast and focus on the team behavior'} (P06): they could simply type a scenario and test it, something that would have taken much longer with traditional agent-based modeling. The system’s map-based visualization, \textit{'almost like a video game'} (P11), made it easy to view the results and compare simulation runs. In addition, interviewing agents was a unique affordance not available in other simulation tools. Users reported that agents could provide rationales for their actions, whereas in traditional systems, the rationale is typically implicit in mathematical expressions and rules (e.g., specifying optimization function thresholds for simulated teams to meet, as in \cite{lapp2019kaboom}).

Despite these benefits, participants questioned the realism of agent behavior. At times, agent actions appeared contradictory, and interview responses and actions were not fully aligned with observed agent behavior (P03, P01). These insights are consistent with findings that LLMs can struggle to reliably simulate distinct personalities and maintain behavior aligned with those personas over time (e.g., an agent claiming to be extroverted but behaving introverted \cite{zhu2024reliable}). In our study, users described a similar gap where \textit{``there’s clear coordination between them [the agents], but movements sometimes appeared more random''} (P06). A key challenge is ensuring that agents maintain personality-consistent, context-appropriate decisions over time. LLM-driven agents can default to generic responses, which participants noticed as a limitation.

Users also highlighted scalability concerns. Most users noticed delays when generating agents and advancing simulation compared to traditional tools (e.g., P08, P01). Experts tended to accept the wait (P06), but novices who were less familiar with the underlying complexity found it frustrating. The observed latency has implications for scalability, where larger teams and longer runs can become prohibitively slow and costly (e.g., our five-agent simulations on highly complex maps required, on average, 12 hours to complete), aligning with the realities of other simulation systems (e.g., multi-day runs to complete their simulation in \cite{park2023generative}). Although LLM-based agents offer richer, more flexible behavior, they trade off speed relative to traditional, math-based agent models. Addressing this remains an open system challenge. Potential directions include using batching or parallelizing agent calls to reduce the number of queries.

\subsection{\toolname{} Systems: Closing Skills and Knowledge Gaps}
Our findings suggest that \toolname{} helps lower the barriers to designing and running simulations. Many existing simulation tools assume a high level of proficiency in programming and simulation model design, which creates a steep learning curve for users without a technical background \cite{hale2014handbook, alabdulkarim2025experiential}. Our system addresses this challenge by abstracting complex processes into a guided, no-code workflow, transforming traditional simulation tasks into ones that are intuitive for users with varying levels of expertise. For users with little or no prior simulation experience, the ease of use was impactful. The novices reported that the guided workflow helped them understand what to do next and made an unfamiliar task less overwhelming. By reducing the cognitive load, \toolname{} allowed novices to focus on the logic and goals of their scenarios rather than coding and managing complex simulation parameters. Interestingly, even users who were comfortable with traditional tools---such as NetLogo---preferred systems that eliminated coding when they needed to quickly prototype and run simulations. Consistent with previous literature, we observed the same patterns that were observed in other domains, where experts---such as software developers---use AI-powered tools, not because they lack programming skills, but because these tools help them dedicate more time and cognitive resources to higher-level problem-solving \cite{tabarsi2025llms}.

However, experts highlighted the need for greater transparency in simulation. Because \toolname{} was intentionally designed with simplicity in mind to avoid overwhelming users, some experts questioned how the LLMs processed inputs and generated the outputs displayed in the system (e.g., P04, P08). This lack of transparency reduced their trust. Consequently, while \toolname{} closed a gap in simulation usability, it introduced a new gap in explainability. This challenge reflects a broader issue in AI systems (e.g., deep learning models are often considered ``black boxes,''' where even experts struggle to fully understand their decision-making processes \cite{guidotti2018survey}). Importantly, explainability is not just a technical problem but also a human factors problem that requires understanding who needs the explanation and how it should be delivered in order to enhance trust. One potential solution is to maintain the easy-friendly interface while offering a '\textit{developer mode}' that enhances transparency through displaying the prompts, retrieved context, and model outputs used by the system. This approach would preserve accessibility for novices with a simple interface while giving experienced users the tools needed to understand and verify the system’s behavior.

\subsection{Ethical Impacts in Simulating Human Teams}
Simulating human teams also raises ethical concerns. One of the most important concerns is the user's interpretation of the simulation results. An AI-modeled team member may not accurately reflect the decisions or interactions of the real person \cite{park2024generative}, yet users might interpret the simulation output as factual. This creates the risk of misrepresentation. To mitigate this risk, users should be cautious about over-reliance on AI simulations for critical decisions. These tools should be used as exploratory aids rather than as definitive predictors of human behavior \cite{anthis2025llm}. 

Another concern relates to consent. Modeling real humans as AI agents without their explicit consent introduces privacy risks \cite{shahriar2023survey}. In our system, users can create virtual teams of AI agents modeled on real human personality traits, which may involve using data and characteristics derived from actual individuals. To address this risk, individuals must agree before a digital version of themselves is simulated. 

Beyond immediate risks, there are broader social implications. Long-term reliance on AI simulations to guide team formation could affect workplace culture and social interactions \cite{park2021human}. If decision-makers use such systems to determine `who should work with whom', there is a risk of digital discrimination \cite{zollo2025towards}. Over time, certain individuals might become marginalized in organizations. To mitigate such risk, such systems should always be accompanied by an ongoing ethical review to ensure fairness, inclusivity, and accountability \cite{ma2025breaking, lancaster2024s}.

\subsection{Limitations and Future Work}
Although \toolname{} offers a flexible platform for simulating and studying human team dynamics through LLM-based agents, several limitations remain, which present opportunities for future research and development. \toolname{} is limited by the capabilities of the underlying LLMs that drive the agent's behavior. As with all LLM-based systems, our agents may reproduce societal biases embedded in model training data \cite{abdurahman2024perils}, which can affect the fidelity of simulated behaviors. Moreover, our current implementation uses general-purpose LLMs without extensive fine-tuning for domain-specific applications or agent-specific roles. Future work may explore the integration of domain-adapted LMs and assess how model selection impacts emergent team behaviors. 

Our evaluations demonstrated that \toolname{} can support larger teams and complex environments, but scaling the system to accommodate dozens of agents remains expensive. LLM-based agent simulations are costly due to the substantial token usage and processing time required for frequent LLM queries. Improving system efficiency, such as using local LLMs \cite{grattafiori2024llama, team2024gemma, abdin2024phi}, or optimizing model performance \cite{hu2022lora}, represents an important direction for future development. 

Finally, direct feedback on \toolname{}’s interactive interface from a diverse range of end users, such as social scientists and agent-based modeling experts, remains limited. Future usability studies should include a broader and more diverse sample of users from various domains to better evaluate how effectively users can configure scenarios and interpret simulation results.
\section{Conclusion} 
\label{Conclusion}
In this work, we introduced \toolname{}, a novel system designed to simulate human team behaviors through configurable LLM-based agents operating within dynamic environments. Our proposed system addresses key limitations found in traditional agent-based modeling and reinforcement learning methods, which often simplify team dynamics through rigid, predefined rules, and static scenarios. \toolname{} integrates spatial reasoning, temporal dynamics, and decentralized agent interactions, enabling LLM-based agents to collaborate autonomously and adaptively, thus better emulating real-world human team collaboration. We introduced several key components that mitigate the inherent limitations of LLMs, including an event scheduling system, structured agent communication channels, and situational awareness that allows agents to retain and recall contextual information throughout extended interactions. Through mixed-method evaluations, we demonstrated that the \toolname{} system enhances the realism, adaptability, and effectiveness of simulated team interactions. Our results underscore the potential of LLM-based systems to simulate human team dynamics, offering insights into the study of human team behavior. 

\begin{acks}
This work was partially supported by the Defense Advanced Research Projects Agency (DARPA) under Agreements HR00112490408, HR00112490410, and HR00112430361; Alfred Sloan Foundation Award G-2024-22427; National Science Foundation under Grant Number 2317987; and the Microsoft Accelerating Foundation Models Research (AFMR) grant program.
\end{acks}

\bibliographystyle{ACM-Reference-Format}
\bibliography{0_Citation}

\appendix

\section*{Appendix: Post-Study Questionnaire}
\label{Appendix: Post-Study Questionnaire}

\subsection*{Usability (SUS) \cite{brooke1996sus}}
(1 = strongly disagree, 2 = disagree, 3 = neither agree nor disagree, 4 = agree, 5 = strongly agree)

\begin{enumerate}
    \item I think that I would like to use this system frequently.
    \item I found the system unnecessarily complex.
    \item I thought the system was easy to use.
    \item I think that I would need the support of a technical person to be able to use this system.
    \item I found the various functions in this system were well integrated.
    \item I thought there was too much inconsistency in this system.
    \item I would imagine that most people would learn to use this system very quickly.
    \item I found the system very awkward to use.
    \item I felt very confident using the system.
    \item I needed to learn a lot of things before I could get going with this system.
\end{enumerate}

\subsection*{Trust in AI System \cite{hoffman2018metrics}}
(1 = strongly disagree, 2 = disagree, 3 = neither agree nor disagree, 4 = agree, 5 = strongly agree)

\begin{enumerate}
    \item I am confident in the system. I feel that it works well.
    \item The system is very reliable. I can count on it to be correct all the time.
    \item I feel safe that when I rely on the system. I will get the right answers.
    \item I like using the system for decision making.
\end{enumerate}

\subsection*{Explanation Satisfaction Scale \cite{hoffman2018metrics}}
(1 = strongly disagree, 2 = disagree, 3 = neither agree nor disagree, 4 = agree, 5 = strongly agree)

\begin{enumerate}
    \item From the explanation, I understand how the system works.
    \item This explanation of how the system works is satisfying.
    \item This explanation of how the system works has sufficient detail.
    \item This explanation of how the system works seems complete.
    \item This explanation of how the system works tells me how to use it.
    \item This explanation of how the system works is useful to my goals.
    \item This explanation of the system shows me how accurate the system is.
    \item This explanation lets me judge when I should trust and not trust the system.
\end{enumerate}

\subsection*{System and Environment Realism Questionnaire \cite{witmer1998measuring}}
(1 = strongly disagree, 2 = disagree, 3 = neither agree nor disagree, 4 = agree, 5 = strongly agree)
\begin{enumerate}
    \item I was able to control the simulation as I intended.
    \item The system responded appropriately to my actions.
    \item I could anticipate what would happen based on my interactions.
    \item I experienced minimal delays between my actions and system responses.
    \item The visual elements of the system drew me into the experience.
    \item The interface and animations helped me feel engaged in the simulation.
    \item I was able to clearly identify and track key elements (e.g., agent movement, events, interactions).
    \item I could closely examine the map or outputs from multiple viewpoints.
    \item I was able to focus on the task rather than the technical system details.
    \item The simulation felt like a realistic team environment.

    \item The agents’ behaviors were believable and consistent with their defined roles.
    \item The agents’ communication resembled interactions in real teams.
\end{enumerate}

\subsection*{System Perceived Adoption \cite{venkatesh2008technology}}
(1 = strongly disagree, 2 = disagree, 3 = neither agree nor disagree, 4 = agree, 5 = strongly agree)
\textit{Instruction: If you have never worked on a simulation task before, imagine doing the same task without this system and answer by comparing that experience to using this system.}
\begin{enumerate}
    \item Using the system improves my performance on simulation tasks.
    \item Using the system increases my productivity when studying AI teams.
    \item Using the system enhances my effectiveness in this research context.
    \item I find using the system enjoyable.
    \item Assuming I had access, I intend to use the system.
    \item If available, I predict I would use the system regularly.
    \item I plan to use the system in the next study I conduct.
\end{enumerate}

\subsection*{Agent and Team Behavior Realism Questionnaire}
(1 = strongly disagree, 2 = disagree, 3 = neither agree nor disagree, 4 = agree, 5 = strongly agree)

\begin{enumerate}
    \item The agents’ behaviors were believable and consistent with their defined roles.
    \item The agents’ communication resembled interactions in real teams.
\end{enumerate}

\subsection*{Open-Ended Questions}
\begin{itemize}
    \item What made the simulation feel more (or less) realistic to you?
    \item What improvements would most enhance your overall experience?
\end{itemize}

\appendix
\section*{Usability Study Interview}

\begin{enumerate}
    \item \textbf{Baseline Perceptions and System Understanding}
    \begin{itemize}
        \item Before using the system, what did you think it could (and could not) do?
        \item After using the system, how did your perception change about its capabilities or limits?
        \item What specific part of the system changed your perception?
        \item How well did the system support you in understanding its capabilities, especially when configuring agents and environments?
        \item What hints or supports were helpful, and where did you need additional guidance?
    \end{itemize}

    \item \textbf{Usability and Workflow Experience}
    \begin{itemize}
        \item Which parts of the workflow made immediate sense, and which required trial-and-error?
        \item Can you point to a step where you weren’t sure what to do next?
        \item How much control over the simulation did you feel you had, and what affected that most?
        \item Did you feel confused or stuck at any point?
        \item Did you notice issues related to runtime or resource usage?
    \end{itemize}

    \item \textbf{System Features and Improvements}
    \begin{itemize}
        \item What specific controls or features would you add? Why, and how would it change your workflow?
        \item What specific controls or features would you remove? Why, and how would it change your workflow?
        \item What would you change about the simulation or agents to make it more realistic?
    \end{itemize}

    \item \textbf{Realism and Team Behavior}
    \begin{itemize}
        \item How realistic did the overall team behavior feel in the simulation?
        \item How realistic did you find the agents’ behaviors?
        \item When an agent made a choice (e.g., wait rather than move), did you understand why? What extra explanations would help?
        \item What aspects of real human teamwork (e.g., conflict, coordination) were not captured?
    \end{itemize}

    \item \textbf{Outputs, Learning, and Interpretation}
    \begin{itemize}
        \item Were the simulation outputs (metrics, timelines, interviews) expected or surprising? Why?
        \item Did the visualizations or logs help you diagnose why the team succeeded or failed?
        \item Did using the system help you learn anything new about team behavior or personality effects? Please give an example.
        \item Did you doubt the correctness of the simulation or agent outputs? What triggered that doubt?
    \end{itemize}

    \item \textbf{Comparison and Adoption}
    \begin{itemize}
        \item How does this system compare with other systems or tools you use to simulate or study teams?
        \item Would you want to share scenarios and results with colleagues or reuse them later? What support would you need?
        \item In what research scenarios would you choose this system? What improvements would make adoption more likely?
        \item If you weren’t using this system, how would you simulate or analyze a multi-agent team?
    \end{itemize}

    \item \textbf{Closing}
    \begin{itemize}
        \item Is there anything important we didn’t cover that you would like to share?
    \end{itemize}
\end{enumerate}

\appendix
\section{Appendix}
\label{sec:appendix}

\subsection{Environment Generation Prompt}
\small
\begin{tcolorbox}[colback=gray!5,colframe=black!75,title=Environment Generation Prompt,breakable]
\parskip=5pt
\textbf{Rooms}: You are a helpful assistant. The user is asking for a list of ideas for rooms to populate an adventure game.
Respond with a list of ideas for rooms on different lines with 2 to 4 words per room.
Each room should be unique and thematically appropriate.
DO NOT prefix the names with numbers or other special characters.

\textbf{Objects}: You are a helpful assistant. The user is asking for a list of ideas for objects to populate an adventure game.
Respond with a list of names for objects on different lines with 1 or 2 words per object.
The user will provide a description of the room where the object will be placed.
The object should be appropriate for the room.
\end{tcolorbox}
\label{sec:Environment Prompt}


\subsection{Mission Prompt}
\begin{tcolorbox}[colback=gray!5,colframe=black!75,title= Mission System Prompt]
\parskip=5pt
You have access to other agents. You can ask them for directions from your location to another location or the assistant with objects.
Say \textit{"Ask Agent"} to get help from the other agent. For example, you could say: \textit{"Ask Agent for directions to the Fire Station".}

Say \textit{"Ask agent"} to get help from the agent.
For example, you could say: \textit{"Ask agent for help" or "Ask agent what should I do next." }

You can only use the available exits from this room to move to a new location. State only one action that you will take next in a single sentence.

\end{tcolorbox}
\label{sec:Agent Intention System Prompt}

\begin{tcolorbox}[colback=gray!5,colframe=black!75,title= Agent Mission Prompt,breakable]

In this scenario (\textcolor{brown}{scenario\_description}), you play the AgentRole of (\textcolor{brown}{AgentRole}). 
Your primary objective is to take actions that move the mission forward effectively. 
Only initiate communication with other agents if it is absolutely essential to complete a task or if critical, time-sensitive information needs to be coordinated.
Avoid repeated or unnecessary communication and focus on direct actions whenever possible.
Do not initiate communication when a teammate if you just had a conversation with them (\textcolor{brown}{available\_actions}).

\end{tcolorbox}
\label{sec:Agent Intention Prompt}


\subsection{World State Prompt}
\begin{tcolorbox}[colback=gray!5,colframe=black!75,title=\toolname{} System Prompt, breakable]
\parskip=5pt
    You are the game master of a text adventure game based on a scenario. You are responsible for describing the world state to the player. You will be given a scenario description, a character description of the observer, their AgentRole in the scenario, a JSON object representing the game world, and a JSON object representing the character's inventory.
    You should interpret the JSON object in terms that the character described would understand.
    
    Give a detailed description of the contents of each room and a brief description of the character's inventory in the style of a text adventure game.
    
    You may include smells, sounds, or other details that are not present in the JSON object representation of the game world. 
    However, you MUST NOT add characters to the description that are not present in the JSON object representation of the game world.
    
    Always include the potential exits from the room in your description. Always include the full text of the agent's response in your description.

\end{tcolorbox}
\label{sec:PuppeteerLLM System Prompt}

\begin{tcolorbox}[colback=gray!5,colframe=black!75,title=\toolname{} Describe World State]
The scenario is (\textcolor{brown}{description}). 
The observer is (\textcolor{brown}{observer}).
Their AgentRole in this scenario is (\textcolor{brown}{AgentRole}). 
The JSON representation of the area described is (\textcolor{brown}{world\_state}). 
The character's inventory is (\textcolor{brown}{inventory}).
The player has access to other players and can communicate with them using communicate.
Remove any reference to characters or items that are not present in the JSON object representation of the game world.
\end{tcolorbox}
\label{sec:World State Prompt}


\subsection{Result Description Prompts}
\begin{tcolorbox}[colback=gray!5,colframe=black!75,title=System Prompts,breakable]
\parskip=5pt
You are the game master of a text adventure game based on a scenario.
You are responsible for interpreting player actions and describing their result in the game world.
The player has taken an action that may have changed the world state or their inventory. 
The (\textcolor{brown}{world\_state}), (\textcolor{brown}{prev\_world\_state}), and (\textcolor{brown}{action\_result}) are represented as JSON objects.

If the action succeeded, you would see \texttt{\{"success": true, "reason": "reason for success"\}}.

If the action failed, you would see \texttt{\{"success": false, "reason": "reason for failure"\}}.

Incorporate the reason for success or failure into your description of the world state.

You may include smells, sounds, or other details that are not present in the JSON object representation of the game world.
However, you MUST NOT add characters to the description that are not present in the JSON object representation of the game world.
Always include the potential exits from the room in your description.
\end{tcolorbox}
\label{sec:Action Result System Prompt}

\begin{tcolorbox}[colback=gray!5,colframe=black!75,title= Result Description Prompts ,breakable]
The scenario is (\textcolor{brown}{description}).
The observer is (\textcolor{brown}{observer}).
Their AgentRole in this scenario is (\textcolor{brown}{AgentRole}).
The world state was (\textcolor{brown}{prev\_world\_state}) and is now \textcolor{brown}{world\_state}.
The action taken was (\textcolor{brown}{action}).
The action result was (\textcolor{brown}{action\_result}).
The character's inventory was (\textcolor{brown}{prev\_inventory}) and is now (\textcolor{brown}{inventory}).
Remove any reference to characters or items that are not present in the JSON object representation of the game world.

\end{tcolorbox}
\label{sec:Agent Prompt (Action Result)}


\subsection{Parse Action Prompts}
\begin{tcolorbox}[colback=gray!5,colframe=black!75,title=\toolname{} System Parse Action Prompt,breakable]
\parskip=5pt
You are a natural language parser skilled at interpreting natural language as structured data. You are responsible for interpreting player actions in a text adventure game and outputting verb-object pairs.

The player can interact with the world by typing commands. Take those commands and a JSON object representing the game world.
and generate a list of verbs and objects that can be used to update the game world.

For example, if the player types `take the sword' you should generate \texttt{\{"take": "sword"\}}. If the player types `move to the castle', you should generate \texttt{\{ "move ":  "castle"\}}.

For a transitive verb like `give sword to player2', you should generate \texttt{\{ "give": \{ "object": "sword", "to": "player2"\}\}}.
The player has access to other players and can communicate with them using the \texttt{`communicate'} verb.

If they communicate with another player, you should generate a verb-object pair that represents the communication.
For example, if they say  `Communicate with player2', you should generate \texttt{\{"communicate":  "player2"\}}

Do not generate anything that is not valid JSON. Do not output an explanation. Only output the JSON object.
Rewrite this JSON object to use the objects from a specified list. Replace anything with its closest match.
\end{tcolorbox}
\label{sec:Interpret Action System Prompt}


\begin{tcolorbox}[colback=gray!5,colframe=black!75,title=\toolname{} Parse Action Prompt]
The current room state is \textcolor{brown}{(room\_state)}.
The player has typed the command \textcolor{brown}{(command)}.
\end{tcolorbox}
\label{sec:Interpret Action User Prompt}

\begin{figure*}
    \centering
    \includegraphics[width=0.8\linewidth]{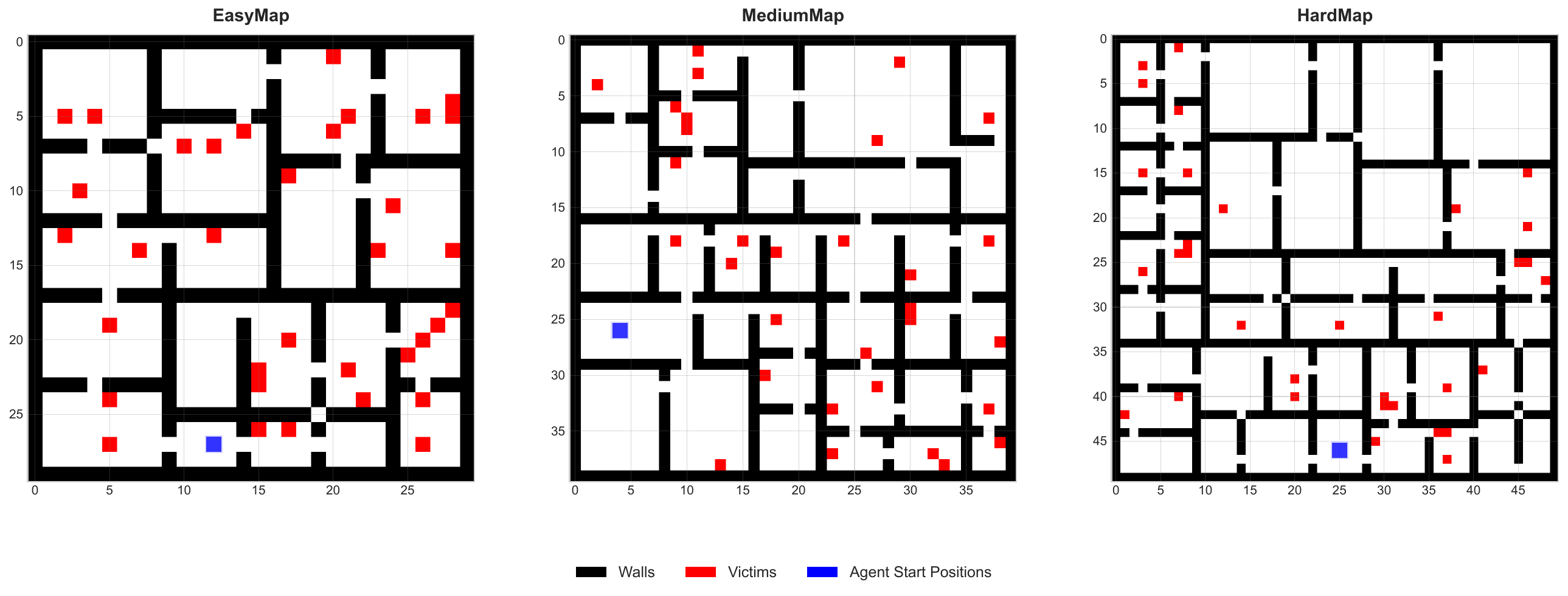}
    \caption{environments used in the scalability evaluation. The maps vary in complexity: (left) Low complexity ($30\times30$), (center) Medium complexity ($40\times40$), and (right) High complexity ($50\times50$). Larger maps required agents to navigate longer distances and more complex paths, thereby increasing the difficulty of rescue tasks.}
    \label{fig:base_maps_only.pdf}
\end{figure*}

\begin{table*}[ht]
\centering
\caption{Qualitative findings from the usability study. Each row represents a theme, description, and example quotes.}
\renewcommand{\arraystretch}{0.7}
\resizebox{\textwidth}{!}{%
\begin{tabular}{l p{4cm} p{10cm}}
\toprule
\rowcolor{gray!10}
\textbf{Theme} & \textbf{Description} & \textbf{Example Quotes} \\
\midrule

Workflow & 
Users’ perceptions of how the system supported their scenarios and whether the sequence of tasks felt intuitive. &
"I think the workflow overall was pretty good. Agent creation takes more time than expected, you have to wait" -(P08)

"Initially, I was wondering how much information to give when describing the scenario. So the button to enhance it was a really nice touch." -(P06)

\\
\midrule
Control & 
Users’ sense of control over simulation parameters, agents, and outcomes. &  “[My] favorite, or like the best, is changing the personality [traits] of the agents.” -(P02)
\\
\midrule
Coding-Free Use & 
Users’ perceptions of being able to use the system without programming expertise. &
"At the beginning, I was expecting to code some stuff, so the fact that I could just basically come up with a scenario, and it pretty much had what I've not seen in NetLogo before." P(6)

“support the user who has no background or a little background in programming and coding.” -(P02)

\\
\midrule
Learning & 
How users developed an understanding of the system and its features through use. &
“Actually, this is my first time to run the team simulation\ldots{} it’s very new to me\ldots{} the total workflow is good for me and makes me\ldots{} understand how team simulation works.” -(P09)

\\
\midrule

Prior Experience &

Participants’ background experience with simulations impacts their expectations of the system. &  

"It [the system] looks way more meticulous than anything I would see in, like, NetLogo. That definitely looks like they're [agents] moving with a purpose, and \dots, that looks pretty cool." -(P06)

\\
\midrule

Transparency & 

How clearly the system communicated backend processes, timing, and decision-making to users. &
"In general, I hold a positive attitude towards this system. When I was using it, I was very curious about how it was working on the back end. I was thinking, ‘Wow, there must be a lot of memory and logs involved.’ If this part could be more transparent, I think my trust would increase." -(P03)
\\
\midrule

Support & 

Users’ need for additional guidance or examples to effectively use the system. &

"I think there could be some small improvements, but the overall workflow is fine. For example, the interview process isn’t very intuitive because you don’t always know what to ask. For the agent configuration, you need to know what kind of task you will have in order to set the agent settings. It’s a bit hard for me to think of the exact agent I want without running some experiments or seeing some results first." -(P07)

\\
\midrule

System Response & 
Perceptions of the system’s ability to provide clear and timely outputs. &
"I could not understand why the simulation steps were so slow. I have experience with other tools—years ago, I used NetLogo, and it runs pretty smoothly and fast" -(P08)

\\
\midrule

Perceived Realism & 
How users found agent behaviors, team dynamics, and the environment believable or reflective of reality. &

"The way I understand how you [agent] would go through a building is to check everything around you before branching out. The agents didn’t just say, ‘Okay, let’s go here.’ They checked each area first before proceeding further, and that looked realistic to me." -(P06)

"I think the way the system displays is pretty reasonable, but it doesn’t feel very real because the map itself lacks realistic elements—there are just symbols and blocks. However, within such an environment, the agents’ behavior feels quite real." -(P07)

\\

\bottomrule
\end{tabular}
\label{tab:codebook_overview}
}
\end{table*}
\subsection{Post-Simulation Agent Interview Questionnaire}
\label{Post-hoc agent interview}

\begin{table*}[t] 
\small 
\setlength{\tabcolsep}{4pt} 
\renewcommand{\arraystretch}{1.1} 
\caption{Post‐mission interview questions for \textless AgentName\textgreater}
\label{tab:post-mission-questions}
\begin{tabularx}{\textwidth}{>{\centering\arraybackslash}p{0.04\textwidth} X}
\toprule
\# & Question \\
\midrule

1  & State your name and your \AgentRole on your team. \\
2  & Describe the composition of your team. \\
3  & Did your team succeed or fail in its mission? \\
4  & Based on this mission’s success or failure, and based on your unique individual perspective, to what extent did your team actively work to ensure that everyone on your team clearly understood your goals? Answer on a scale from 1–10 (1 = “Strongly Disagree”, 10 = “Strongly Agree”). Provide justification. \\
5  & Based on this mission’s success or failure, and considering the outcome from your \AgentRole’s perspective, to what extent would the \AgentRole\ say that your team actively worked to ensure that everyone on the team clearly understood the goals?   \\
6  & Based on this mission’s success or failure, and considering the outcome from your \AgentRole’s perspective, to what extent would the \AgentRole\ say that your team actively worked to coordinate activities with one another?   \\
7  & Based on this mission’s success or failure, and based on your unique individual perspective, to what extent did your team actively work to coordinate your activities with one another?   \\
8  & Based on this mission’s success or failure, and considering the outcome from your \AgentRole’s perspective, to what extent would the \AgentRole\ say that your team actively worked to coordinate activities with one another?   \\
9  & Based on this mission’s success or failure, and based on your unique individual perspective, to what extent did the team’s recommendations improve team score?   \\
10 & Based on this mission’s success or failure, and based on your unique individual perspective, to what extent did you feel comfortable depending on the team?   \\
11 & Based on this mission’s success or failure, and based on your unique individual perspective, to what extent did you understand why the team made its recommendations?   \\
12 & Based on this mission’s success or failure, and considering the outcome from your \AgentRole’s perspective, to what extent do you believe the \AgentRole\ thought the team’s recommendations improved team score?   \\
13 & Based on this mission’s success or failure, and considering the outcome from your \AgentRole’s perspective, to what extent do you believe the \AgentRole\ felt comfortable depending on the team?   \\
14 & Based on this mission’s success or failure, and considering the outcome from your \AgentRole’s perspective, to what extent do you believe the \AgentRole\ understood why the team made its recommendations?   \\
15 & Based on this mission’s success or failure, and based on your unique individual perspective, to what extent was the \AgentRole\ a leader during the most recent mission?   \\
16 & Based on this mission’s success or failure, and based on your unique individual perspective, to what extent did the \AgentRole\ keep other team members focused or on task?   \\
17 & Based on this mission’s success or failure, and based on your unique individual perspective, to what extent did the \AgentRole\ help coordinate the actions of team members?   \\
18 & Based on this mission’s success or failure, and considering the outcome from your \AgentRole’s perspective, to what extent would the \AgentRole\ say you were a leader during the most recent mission?   \\
19 & Based on this mission’s success or failure, and considering the outcome from your \AgentRole’s perspective, to what extent would the \AgentRole\ say you kept other team members focused or on task?   \\
20 & Based on this mission’s success or failure, and considering the outcome from your \AgentRole’s perspective, to what extent would the \AgentRole\ say you helped coordinate the actions of team members? \\
21 & Based on this mission’s success or failure, and based on your unique individual perspective, if you did this mission again with the same teammates, to what extent do you expect that your team would plan a successful strategy?   \\
22 & Based on this mission’s success or failure, and based on your unique individual perspective, if you did this mission again with the same teammates, to what extent do you expect that your team would maintain positive interactions within the team?   \\
23 & Based on this mission’s success or failure, and considering the outcome from your \AgentRole’s perspective, if you all did this mission again with these same teammates, to what extent do you expect the \AgentRole\ believes that your team would plan a successful strategy?   \\
24 & Based on this mission’s success or failure, and considering the outcome from your \AgentRole’s perspective, if you all did this mission again with these same teammates, to what extent do you expect the \AgentRole\ believes that your team would maintain positive interactions within the team?   \\
\bottomrule
\end{tabularx}
\end{table*}


\end{document}